\documentclass[%
reprint,
amsmath,amssymb,
aps,
]{revtex4-1}

\usepackage[pdftex]{hyperref} 
\usepackage{graphicx}
\usepackage{dcolumn}
\usepackage{bm}

\begin{document}


\title{Building traps for skyrmions by the incorporation of magnetic defects into nanomagnets: pinning and scattering traps by magnetic properties engineering}

\author{D.~Toscano}\email[Author to whom correspondence should be addressed. Electronic mail: ]{danilotoscano@fisica.ufjf.br}
\author{S.~A.~Leonel}
\author{P.~Z.~Coura} 
\author{F.~Sato} 

\affiliation{Departamento de F\'{\i}sica, Laborat\'orio de Simula\c{c}\~ao Computacional, Universidade Federal de Juiz de Fora, Juiz de Fora, Minas Gerais 36036-330, Brazil} 

%
%

\date{\today}

\begin{abstract}
\noindent{In this work we have used micromagnetic simulations to report four ways to build traps for magnetic skyrmions. Magnetic defects have been modeled as local variations in the material parameters, such as the exchange stiffness, saturation magnetization, magnetocrystalline anisotropy and  Dzyaloshinskii-Moriya constant. We observe both pinning (potential well) and scattering (potential barrier) traps when tuning either a local increase or a local reduction for each one of these magnetic properties. It is found that the skyrmion-defect aspect ratio is a crucial parameter to build traps for skyrmions. In particular, the efficiency of the trap is compromised if the defect size is smaller than the skyrmion size, because they interact weakly. On the other hand, if the defect size is larger than the skyrmion diameter, the skyrmion-defect interaction becomes evident. Thus, the strength of the skyrmion-defect interaction can be tuned by the modification of the magnetic properties within a region with suitable size. Furthermore, the basic physics behind the mechanisms for pinning and for scattering is discussed. In particular, we discover that skyrmions move towards the  magnetic region which tends to maximize its diameter; it enables the magnetic system to minimize its energy. Thus, we are able to explain why skyrmions are either attracted or repelled by a region with modified magnetic properties. Results here presented  are of utmost significance for the development and realization of future spintronic devices, in which skyrmions will work as information carriers.}
\end{abstract}

\maketitle


%
\section{\label{sec:Intro}Introduction}
Topologically protected objects known as skyrmions were first introduced by Skyrme in the context of particle physics~\cite{Proc_R_Soc_London_Ser_A_260_127_1961,Nucl_Phys_31_556_1962}. From then on, the concept of these topological excitations has been extended to the condensed matter physics, where skyrmions can be found in liquid crystals~\cite{Rev_Mod_Phys_61_385_1989,Nat_Commun_2_246_2011}, Bose-Einstein condensates~\cite{PhysRevLett_81_742_1998,Nat_London_411_918_2001,New_J_Phys_17_069501_2015}, superconductors~\cite{PhysRevLett_119_167001_2017,JPhys_CondensMatter_30_295601_2018} and nanoscaled magnetic thin films~\cite{Belavin_Polyakov_1975,JMagnMagnMater_138_255_1994,JMagnMagnMater_305_413_2006,Nature_442_797_2006,NatureNanoTechnology_8_839_2013,NatureNanoTechnology_8_899_2013}. 

Nanomagnets are suitable systems to investigate exotic magnetic structures, such as vortices, skyrmions and domain walls. These quasiparticles are not only relevant to fundamental micromagnetism, but also enable us to engineer spintronic devices~\cite{Nat_Mater_6_813_2007,Nat_Rev_Mater_2_17031_2017}. In this work we focus on skyrmions, the information carriers that will probably form the basis of the next generation in data storage and logic devices~\cite{JPhysD_ApplPhys_44_392001_2011,NatureNanoTechnology_8_152_2013}. Initially, magnetic skyrmions has been experimentally observed at low temperature and under external magnetic fields~\cite{Science_323_915_2009,PhysRevB_81_041203_2010,Nature_465_901_2010,NaturePhysics_7_713_2011,Nat_Mater_10_106_2011,Science_336_198_2012,
Science_341_636_2013}. Nowadays, they have been stabilized at room temperature without the aid of a magnetic field~\cite{Science_349_283_2015,ApplPhysLett_106_242404_2015,Nat_Mater_15_501_2016,NatureNanoTechnology_11_444_2016,NatureNanoTechnology_11_449_2016,
ApplPhysLett_111_202403_2017,ApplPhysLett_112_132405_2018}. 
Usually skyrmions arise in magnetically ordered systems with Dzyaloshinskii-Moriya (DM) couplings~\cite{Dzyaloshinskii_J_Phys_Chem_Solids_4_241_1958,Moriya_PhysRev_120_91_1960,JMagnMagnMater_182_341_1998,
PhysRevLett_87_037203_2001,JPhys_CondensMatter_24_086001_2012}, which can be either intrinsic in chiral magnets (for example, compounds of transition metals  with the noncentrosymmetric B20-type structure) or induced in magnetic multilayer systems with broken inversion symmetry and strong spin-orbit coupling (for example, in Co/Pt multilayers). Indeed, skyrmions can be stabilized in the absence of DM interactions~\cite{NatureNanoTechnology_8_899_2013,PhysRevLett_105_197202_2010,PhysRevB_88_054403_2013}, but it generates giant skyrmions. In these magnetic systems, the skyrmion sizes are typically of the same order of magnitude than the vortex sizes (from 100 nm to 1 $\mu$m). On the other hand, in magnetic systems in which DM interactions are significant, the skyrmion diameters are of the order of tens of nanometers. Unlike vortices which have a free chirality (clockwise or anticlockwise rotation), skyrmions have a fixed rotation sense which is specified by the sign of the DM interaction parameter. Importantly, several works refer to this structural property of the skyrmion as helicity instead of chirality~\cite{NatureNanoTechnology_8_899_2013,NatureNanoTechnology_8_723_2013}. 
Furthermore, skyrmions can occur in two types~\cite{NatureNanoTechnology_8_152_2013}: Bloch's skyrmion (vortex-type configuration) and N\'{e}el's  skyrmion (hedgehog-type configuration). Thus, magnetically ordered systems with DM interactions can exhibit either a single skyrmion or a close-packed lattice of identical skyrmions.



Due to the peculiar characteristics of the skyrmions, nanoscale sizes, topological protection, and low current densities needed to drive them, there is a lot of interest in replacing domain walls with skyrmions in spintronic technologies. The overview as well as a sketch of skyrmionic racetrack is presented in Ref.~\cite{Science_349_234_2015}. The stability of skyrmions has been extensively investigated and there are many ways of writing and deleting them in magnetic nanostructures~\cite{NatureNanoTechnology_8_839_2013,NatureNanoTechnology_8_723_2013,Science_349_283_2015,
PhysRevB_85_174416_2012,ApplPhysLett_102_222405_2013,PhysRevB_88_184422_2013,NatureNanoTechnology_8_742_2013,Nat_Commun_5_4652_2014,
PhysRevLett_114_177203_2015,JPhysD_ApplPhys_48_115004_2015,PhysRevLett_110_167201_2013,Scientific_Reports_5_17137_2015,
AIP_Advances_5_047141_2015,PhysRevB_93_024415_2016}. Skyrmions can be manipulated by spin-polarized currents applied in-plane as well as perpendicular to the plane of the nanostructure~\cite{NatureNanoTechnology_8_839_2013,NatureNanoTechnology_8_899_2013,Science_330_1648_2010,Nat_Commun_3_988_2012,
Nat_Commun_4_1463_2013,PhysRevLett_110_207202_2013,PhysRevB_89_064412_2014,PhysRevB_89_064425_2014,Scientific_Reports_5_10620_2015}. Unlike domain walls which are restricted to the unidirectional movement along the nanotrack, skyrmions can not be driven by spin-polarized currents without being moved away from the longest axis of the nanotrack. As a result of the skyrmion Hall effect~\cite{NatureNanoTechnology_8_899_2013,NaturePhysics_13_162_2017}, skyrmions can be accumulated or even annihilated at the edges of the nanowire. The reason why the skyrmions move away from the nanotrack longest axis when applying a spin-polarized current can also be understood through Thiele approach~\cite{PhysRevLett_30_230_1973,PhysRevB_14_3130_1976}, which takes into account the Magnus force~\cite{Nat_Commun_4_1463_2013,NaturePhysics_8_301_2012}. From the technological point of view, this is an issue to the development of spintronic devices based on skyrmion transport and various strategies have been planned to suppress this undesirable phenomenon. An experimental challenge is to find a special class of ferromagnetic materials which presents the same value for the Gilbert damping and non-adiabatic spin-transfer torque parameters, i.e., $\alpha=\beta$. In this case, there is no skyrmion Hall effect, as observed in micromagnetic simulations~\cite{NatureNanoTechnology_8_839_2013,PhysRevB_85_174416_2012}. Instead of using electrical currents to control the skyrmion transport, spin waves~\cite{PhysRevLett_108_017601_2012,Nanotechnology_26_225701_2015} as well as temperature gradients~\cite{PhysRevLett_111_067203_2013,PhysRevLett_112_187203_2014} can be used to drive skyrmions in nanotracks. Another possibility for controlling the skyrmion motion along the nanotrack central axis by using a spin-polarized current is to change the magnetic medium, that is, when replacing the ferromagnet with the antiferromagnet~\cite{Scientific_Reports_6_24795_2016,PhysRevB_96_060406_2017}. Furthermore, it was reported that the skyrmions were significantly influenced by defects in the magnetic medium. The inhomogeneity of a material can be intrinsic (impurities) or induced (intentionally incorporated imperfections). Generally, non-magnetic defects, which distort locally the nanomagnet geometry, generate the skyrmion pinning in the neighborhood of a hole, the region with missing magnetic moments~\cite{PhysRevB_89_054434_2014,PhysRevB_91_054410_2015}. Chappert et al.~\cite{Ion_Irradiation_1998} pioneered the located modification of the magnetic properties by employing the ion irradiation in magnetic thin films and multilayers, for a review see Ref.~\cite{Review_Implantation_2008}. As a result, magnetic defects can be intentionally incorporated in nanomagnets in order to create traps for vortex~\cite{ApplPhysLett_101_252402_2012,JMagnMagnMater_324_3083_2012}, domain walls~\cite{JMagnMagnMater_419_37_2016,JApplPhys_116_163901_2014,PhysRevApplied_3_034008_2015} as well as skyrmions~\cite{ApplPhysLett_109_042406_2016}. References~\cite{JApplPhys_116_163901_2014,PhysRevApplied_3_034008_2015} showed that it is possible to use focused ion beam irradiation to modify the saturation magnetization at the interface of non-magnetic/ferromagnetic multilayers. We believe that the intermixing of a non-magnetic metal and Permalloy, induced by the Ga$^{+}$ ion beam at the interface of the multilayer, is able to vary not only the saturation magnetization but also other material parameters in the irradiated area. In particular, the exchange stiffness constant, since the exchange energy represents the main term in the total energy of a ferromagnet, see~\cite{Review_Implantation_2008}. Once DM interactions are induced in ferromagnetic/heavy metal  multilayer systems with broken inversion symmetry and strong spin-orbit coupling (for example, in Co/Pt multilayers), we believe the modification of the DM interaction strength can be achieved by varying of layer thickness in a selected region, see Refs.~\cite{PhysRevLett_115_267210_2015,Nano_Lett_17_2703_2017}.

The influence of a triangular shaped defect characterized by a local increase of the easy axis anisotropy has been extensively investigated using micromagnetic simulations~\cite{NatureNanoTechnology_8_152_2013,NatureNanoTechnology_8_839_2013,JPhysD_ApplPhys_48_115004_2015}, and it was observed that skyrmions are repelled  by this magnetic defect. It is known since 1998 that the simplest effect induced by the irradiation of He$^{+}$ ion in a Co/Pt multilayer is to reduce the perpendicular anisotropy of the irradiated area~\cite{Ion_Irradiation_1998,ApplPhysLett_109_042406_2016}. Not surprisingly, Fook et al.~\cite{IEEE_51_1500204_2015} observed the confinement of skyrmions along of a reduced anisotropy groove strategically located on the axis of the nanotrack. Thus, skyrmions are attracted by a region of reduced perpendicular anisotropy and can be driven by an in-plane spin-polarized current without being drift from the direction of the electron flow. A similar strategy using potential barriers instead of potential wells has been proposed recently~\cite{Scientific_Reports_7_45330_2017}. In particular, the top and bottom edges of the nanotrack were considered made of a material with a higher perpendicular anisotropy. Another kind of magnetic defect, consisting in local variations of the exchange constant, has been investigated in Refs.~\cite{JPhys_CondensMatter_25_076005_2013,PhysRevB_87_214419_2013}. In their micromagnetic simulations, the authors observed pinning center as well as obstacle for magnetic skyrmions. In previous works, our team has modeled magnetic defects as located variations on the exchange stiffness constant and it was observed pinning and scattering traps for both the vortex core~\cite{JAP_109_076104_2011} and the domain wall~\cite{JAP_114_013907_2013}. In summary, we have observed that a local decrease of the exchange constant work as a pinning center, whereas a local increase of the exchange constant work as a scattering center for the quasiparticle in soft ferromagnetic materials, such as Permalloy. In this work, similar mechanisms of pinning (attractive interaction) and scattering (repulsive interaction) have also been observed for skyrmions in ferromagnetic nanowires with perpendicular magnetic anisotropy, provided that the skyrmion size is smaller than defect size.
%
%
Very recently, a study using atomistic spin simulation investigated the interaction of skyrmions with different kinds of atomic-scale defects~\cite{PhysRevB_96_214403_2017}. The authors considered magnetic moments arranged on a triangular lattice, and magnetic defects having a hexagonal geometry with typical size of the order of a few nanometers (from 1 to 5 nm). 
Mechanisms to pin skyrmions in a Co monolayer on Pt substrate were observed for different kinds of magnetic defects: either a local reduction of the exchange constant, a local increase of the strength of the DM interaction, or a local reduction of the magnetic anisotropy.

As we have reviewed, there are a number of works in the literature that have investigated the skyrmion-defect interaction. However, these works have considered defects which are relatively small. Besides ensuring a poor thermal stability around the trap, the incorporation of small defects in magnetic nanostructures can be a challenge for current experimental techniques. Moreover, the mechanisms behind of skyrmion pinning reported in these works are very difficult to understand. Up to now, no work explains why skyrmions are either attracted or repelled by a region modified magnetically. Besides complementing the previous studies by including the effect of the dipolar coupling, our results indicate that it is also possible to
obtain skyrmion traps through located variations in the saturation magnetization.
Our work system consists in Co/Pt multilayer nanowires hosting a single skyrmion with the hedgehog configuration. In such ferromagnetic nanotracks, it was predicted theoretically~\cite{Scientific_Reports_4_6784_2014} and was observed experimentally~\cite{Science_349_283_2015,ApplPhysLett_106_242404_2015,Nat_Mater_15_501_2016,NatureNanoTechnology_11_444_2016,NatureNanoTechnology_11_449_2016,
ApplPhysLett_111_202403_2017,ApplPhysLett_112_132405_2018} a high thermal stability.  
%
%
%
%
%
%
The main goal in this paper is to investigate how skyrmions are influenced by a located variation on the intrinsic parameters of the ferromagnetic nanotrack, including exchange stiffness, saturation magnetization, magnetocrystalline anisotropy and  Dzyaloshinskii-Moriya constant. For this purpose, magnetic defects were strategically located on the nanotrack axis and in the neighborhood of the skyrmion. In order to demonstrate the Physics of the skyrmion-defect interaction, we do not consider magnetic system under the influence of an external agent, such as magnetic fields or spin-polarized currents. Our methodology consists in computing energy differences  by the Hamiltonian of the system. Thus, we obtain both well and barrier potentials, which identify, as quickly as possible, the kind of the interaction that we are dealing with, either attractive  or repulsive. Predictions of our results  are verified solving the Landau-Lifshitz-Gilbert equation  without any external agent. Here, we answer the following question, magnetic defects in nanowires: traps for skyrmions in spintronic technologies?

%
\section{\label{sec:Methods}Modeling and Simulation Details}

To describe the magnetic system we have considered exchange, Dzyaloshinskii-Moriya, perpendicular magnetic anisotropy and dipole-dipole interactions, included in the following Hamiltonian:


\begin{eqnarray}
H  & = & - \:\sum_{<i,j>}\: J_{ij}\: \left [\:\hat{m}_{i}\cdot\hat{m}_{j}\:\right ] \:\:+{}                                        
                    \nonumber \\
                    \nonumber \\
&& {}  +  \sum_{<i,j>}\: D_{ij}\: \left [ \:\hat{d}_{ij}\cdot (\hat{m}_{i}\times \hat{m}_{j})\: \right ] \:\:+{}
                    \nonumber \\
                    \nonumber \\
&& {}  - \:\sum_{i}\:  K_{i} \left [\: \hat{m}_{i}\cdot\hat{n}\: \right ]^{2} \:\: +{}
                    \nonumber \\
&& {}  +  \:\sum_{i,j}\: M_{ij} \: \left[\frac{\hat{m}_{i}\cdot\hat{m}_{j} - 3(\hat{m}_{i}\cdot\hat{r}_{ij})(\hat{m}_j\cdot\hat{r}_{ij})}{(r_{ij}/a)^{3}}\right]  
 \label{hamiltonian}
\end{eqnarray}
where $ \hat m_{k} \equiv (m^{x}_{k},m^{y}_{k},m^{z}_{k})$ is a dimensionless vector, corresponding to the magnetic moment located at the site $k$ of the lattice. Once $J_{ij}>0$, the first term in Eq. (\ref{hamiltonian}) describes the ferromagnetic coupling.  Due to the short range of the exchange interaction, the summation is over the nearest magnetic moment pairs $<i,j>$. The second term in Eq. (\ref{hamiltonian}) represents the Dzyaloshinskii-Moriya interactions, where the versor $\hat{d}_{ij}$ depends on the type of magnetic system considered. Dzyaloshinskii-Moriya interactions originate from inversion asymmetry and large spin-orbit coupling. In bulk materials with a lack of spatial inversion symmetry (B20 structures), the versor of the Dzyaloshinskii-Moriya interaction is given by $\hat{d}_{ij}=\hat{u}_{ij}$, where $\hat{u}_{ij}$ is unit vector joining the sites $i$ and $j$ in the same layer. In this case, Bloch skyrmions (vortex-type configuration) can be stabilized in the framework of a classical Heisenberg spin model with Dzyaloshinskii-Moriya interaction. On the other hand, for a magnetic multilayer system containing the interface between a magnetic ultrathin layer and a strong spin-orbit coupling adjacent layer, the versor of the Dzyaloshinskii-Moriya interaction is given by $\hat{d}_{ij}=\hat{u}_{ij}\times \hat{z}$, where $\hat{z}$  is a versor perpendicular to the multilayer surface. These magnetic systems favor the nucleation of  N\'{e}el skyrmions (hedgehog-type configuration). The constant of the Dzyaloshinskii-Moriya interaction $D_{ij}$ can be either positive or negative, thus it defines the chirality of the magnetic structure. For a vortex-like skyrmion, $D_{ij}<0$ produces a structure with clockwise rotation, whereas $D_{ij}>0$ produces a structure with anticlockwise rotation. For a hedgehog-like skyrmion, $D_{ij}<0$ produces a structure with inward radial flow, whereas $D_{ij}>0$ produces a structure with outward radial flow. The third term in Eq. (\ref{hamiltonian}) describes the uniaxial magnetocrystalline anisotropy, since $K_{i}>0$ and $\hat{n}=\hat{z}$ is a versor perpendicular to the magnetic layer surface. The last term in Eq. (\ref{hamiltonian}) represents the dipolar coupling; due to the long range of this interaction we consider all the dipole-dipole interactions. The strength of the magnetic interactions, $J_{ij}, \: D_{ij}, \: K_{i}, \: M_{ij}$ have the same dimension (energy unity) and they assume different values depending on the local variation of the material parameters: exchange stiffness constant $A$, Dzyaloshinskii-Moriya constant $D$, magnetocrystalline anisotropy constant $K$  and saturation magnetization constant $M_{\mbox{\tiny{S}}}$. In our simulations we use the micromagnetic approach, in which the work cell has an effective magnetic moment $\vec{m}_{k}=\left(M_{\mbox{\tiny{S}}}\:V_{cell}\right)\:\hat{m}_{k}$. For the case in which the magnetic system is discretized into cubic cells $V_{cell}=a^{3}$, the possible values for the strength of the magnetic interactions are as follow:

\begin{equation}
J_{ij}= 2\: a\: \left\{\begin{array}{l}
         A \\
         A' \\
         A''   
             \end{array}\right.
\label{eq:A}
\end{equation}

\begin{equation}
D_{ij}= a^{2} \left\{\begin{array}{l}
         D \\
         D' \\
         D''   
             \end{array}\right.
\label{eq:D}
\end{equation}
\begin{equation}
K_{i}= a^{3} \left\{\begin{array}{l}
         K \\
         K''   
             \end{array}\right.
\label{eq:K}
\end{equation}


\begin{equation}
M_{ij}= \left ( \frac{\mu_{0}\: a^{3}}{4\pi}  \right ) \left\{\begin{array}{l}
         M_{\mbox{\tiny{S}}}\:\: M_{\mbox{\tiny{S}}}\\
         M_{\mbox{\tiny{S}}}\:\: M_{\mbox{\tiny{S}}}''\\
         M_{\mbox{\tiny{S}}}''\:\: M_{\mbox{\tiny{S}}}''   
             \end{array}\right.
\label{eq:Ms}
\end{equation}
where $A, D, K, M_{\mbox{\tiny{S}}}$ are parameters of the host material. $A'', D'', K'', M_{\mbox{\tiny{S}}}''$ are the parameters of the guest material. $A', D'$ are parameters which describe interactions at the interface between two ferromagnetic materials, see figures \ref{fig:skyrmions_interface}(a) and \ref{fig:Schematic}. The geometric mean is adopted for the interface parameters: $A'=\sqrt{A\:\cdot\:A''\:\:}$ and $D'=\sqrt{D\:\cdot\:D''\:\:}$ in order to allow the magnetic parameters to vary gradually.

In the simulations we chose the typical parameters for Co/Pt multilayers, the values are as follow: exchange stiffness constant $A=1.5 \times 10^{-11}$ J/m, Dzyaloshinskii-Moriya constant $D=4.0\times 10^{-3}$ J/m$^{2}$, magnetocrystalline anisotropy constant $K=1.2\times 10^{6}$ J/m$^{3}$  and saturation magnetization constant $M_{\mbox{\tiny{S}}}=5.8\times 10^{5}$ A/m. In order to choose a suitable size for the work cell we estimate the characteristic lengths that are relevant to the problem: the exchange length $\lambda = \sqrt{\dfrac{2A}{\mu_0 M_{\mbox{\tiny{S}}}^2}}\approx 8.42\: \textrm{nm}$, the wall width parameter $\Delta = \sqrt{\dfrac{A}{K}}\approx 3.54\: \textrm{nm}$ and the length associated with the Dzyaloshinskii-Moriya interaction $\xi= \frac{2A}{D}\approx 7.50\: \textrm{nm}$, see Ref.~\cite{PhysRevB_88_184422_2013}. The size of the work cell used in the simulations was $V_{cell}=2\times 2\times 2\:\textrm{nm}^{3}$. Once the side of the cell is smaller than the smallest characteristic lengths, $a=2\:\textrm{nm}<\Delta$, the chosen unit cell of $2\times 2\times 2\:\textrm{nm}^{3}$ is accurate enough for the current study. For the geometric parameters of the planar nanowires we have considered the length $l=500\:\textrm{nm}$, the width $w=60\:\textrm{nm}$ and the thickness $t=2\:\textrm{nm}$, differing one to another only in the parameters of the magnetic defect: the local variation of the magnetic property into a region of area $S$. Guest material parameters, $A'', D'', K'', M_{\mbox{\tiny{S}}}''$ are regarded as tuning parameters to obtain traps for the skyrmion. We have considered magnetic defects with different areas, where S ranging from 4 to 2116 $\textrm{nm}^{2}$, containing either a local reduction  or a local increase of the magnetic properties. It is worth mentioning that local variations of the magnetic properties were considered individually.
Thus, we have studied four possible sources of magnetic defects: Type $A$, Type $D$, Type $K$ and Type $M_{\mbox{\tiny{S}}}$. Type $A$ magnetic defects are those characterized only by local variations in the exchange stiffness constant (other parameters of the magnetic material were unchanged in the region of the defect), Type $D$ magnetic defects are those characterized only by local variations in the Dzyaloshinskii-Moriya constant, and so on. 

The magnetization dynamics is governed by the Landau-Lifshitz-Gilbert (LLG) equation, whose discrete version can be written as following:

\begin{equation}
 \frac{d\hat{m}_{i}}{d\tau} = -\frac{1}{1+\alpha^{2}} \left[ \hat{m}_{i} \times \vec{b}_{i} \: + \: \alpha \:\: \hat{m}_{i}\times ( \hat{m}_{i} \times  \vec{b}_{i} ) \right]
\label{motion}
\end{equation}
where $\vec{b}_{i}$ is the dimensionless effective field at the lattice site $i$, containing individual contributions from the exchange, Dzyaloshinskii-Moriya, anisotropy, dipolar, and Zeeman fields. The Gilbert damping parameter is fixed at $\alpha=0.3$, which corresponds to a typical value for  Co/Pt multilayers. The connection between the time and its dimensionless corresponding is given by $d\tau=\nu\:dt$, where $\nu=\left(\dfrac{\lambda}{a}\right)^{2}\gamma\:\mu_{0}\:M_{\mbox{\tiny{S}}}$, being $\gamma=1.76\times10^{11}\:\textrm{(T.s)}^{-1}$ the electron gyromagnetic ratio. The LLG equation was integrated by using a fourth-order predictor-corrector scheme with time step $\Delta\tau=0.01$.

In order to obtain the remanent of the planar nanowire with a single skyrmion we have chosen as initial condition an analytical solution in which a N\'{e}el skyrmion is placed exactly at the geometric center of the nanowire. If no external agent (magnetic field or current) were present, the integration of the LLG equation, Eq. (\ref{motion}), leads the magnetic system to the minimum energy configuration. This makes possible the adjustment not only of the skyrmion radius but also its out-of-plane magnetization. The equilibrium configuration obtained in this way has been used as initial configuration in other simulations where a magnetic defect was inserted into the nanowire. To calculate the interaction energy $U(s)$ between the skyrmion and the magnetic defect as a function of the center-to-center separation $s$, we have fixed the skyrmion at the center of the nanowire and only varied the defect position along the nanowire axis, see Fig. (\ref{fig:Schematic}). For each separation $s$, the total energy $E(s)$ of the system was calculated using the Eq. (\ref{hamiltonian}), and the interaction energy has been estimated using the following expression:
\begin{equation}
U(s)=E(s)-E(s \to \infty)
\label{interaction_energy}
\end{equation}

After computing the interaction energy between the skyrmion and the magnetic defect, we investigate how magnetic defects affect the dynamics of the skyrmion when integrating the LLG equation, Eq. (\ref{motion}), with zero external agent. 

%
%

%

%
\section{\label{sec:Results}Results and Discussion}

\subsection{\label{sec:NoDefect}Nanotracks made of a single magnetic material}

\begin{figure}[htb!]
\centering
	\includegraphics[width=8.7cm]{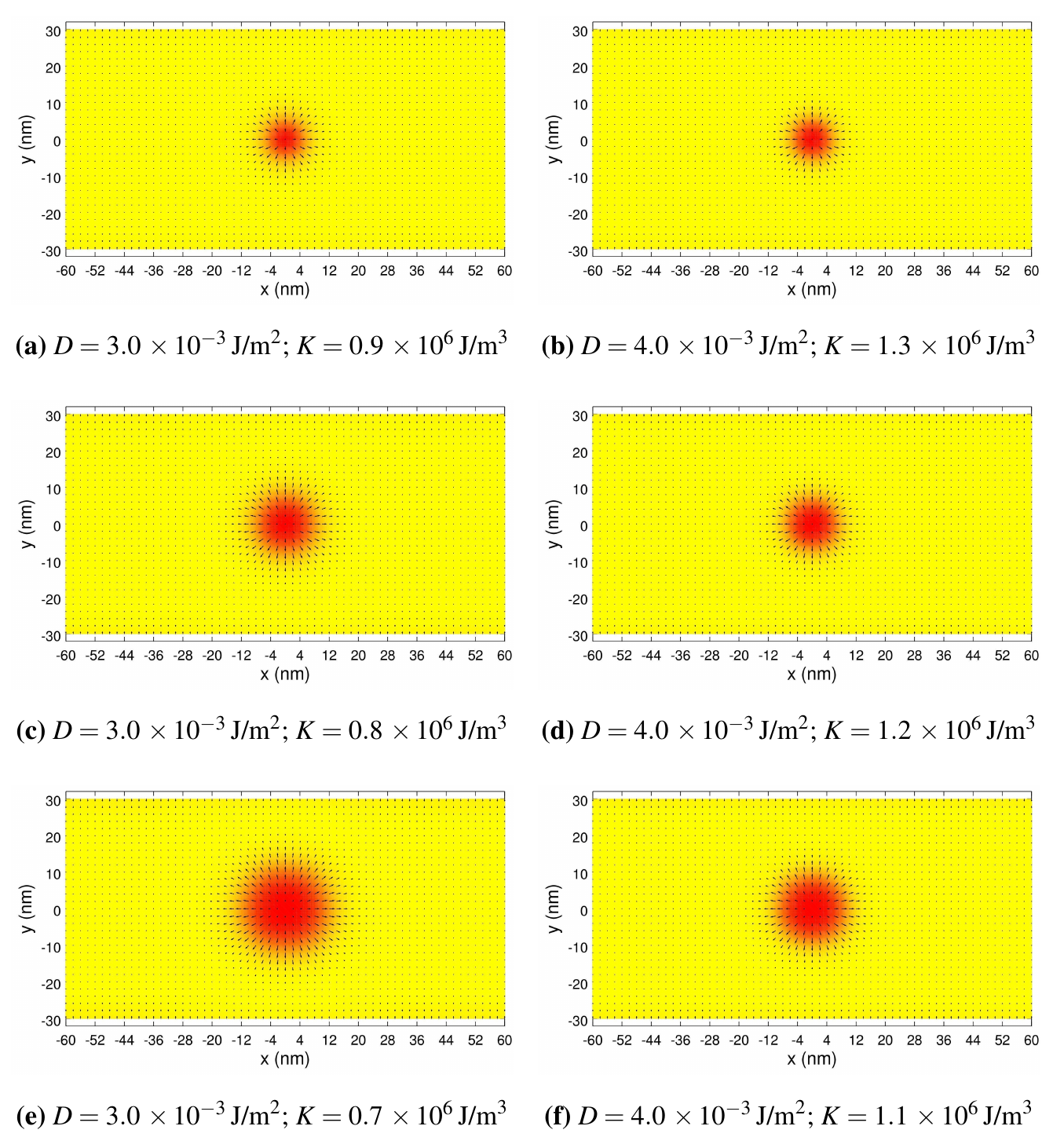}
\caption{(Color online). Equilibrium configurations of a single skyrmion in the planar nanowire 60 nm wide as a function of the magnetic properties of the medium. Figures (a) to (f) were obtained using $A=1.5 \times 10^{-11}$ J/m and $M_{\mbox{\tiny{S}}}=5.8\times 10^{5}$ A/m. In (a) the skyrmion diameter is $d=9.61\:\textrm{nm}$. In (b) the skyrmion diameter is $d=10.01\:\textrm{nm}$. In (c) the skyrmion diameter is $d=15.02\:\textrm{nm}$. In (d) the skyrmion diameter is $d=13.21\:\textrm{nm}$. In (e) the skyrmion diameter is $d=22.02\:\textrm{nm}$. In (f) the skyrmion diameter is $d=17.62\:\textrm{nm}$. Unless otherwise stated, the skyrmion of figure (d) was used in most of our simulations.}
\label{fig:skyrmions_sizes}
\end{figure}

\begin{figure}[htb!]
\centering
	\includegraphics[width=8.7cm]{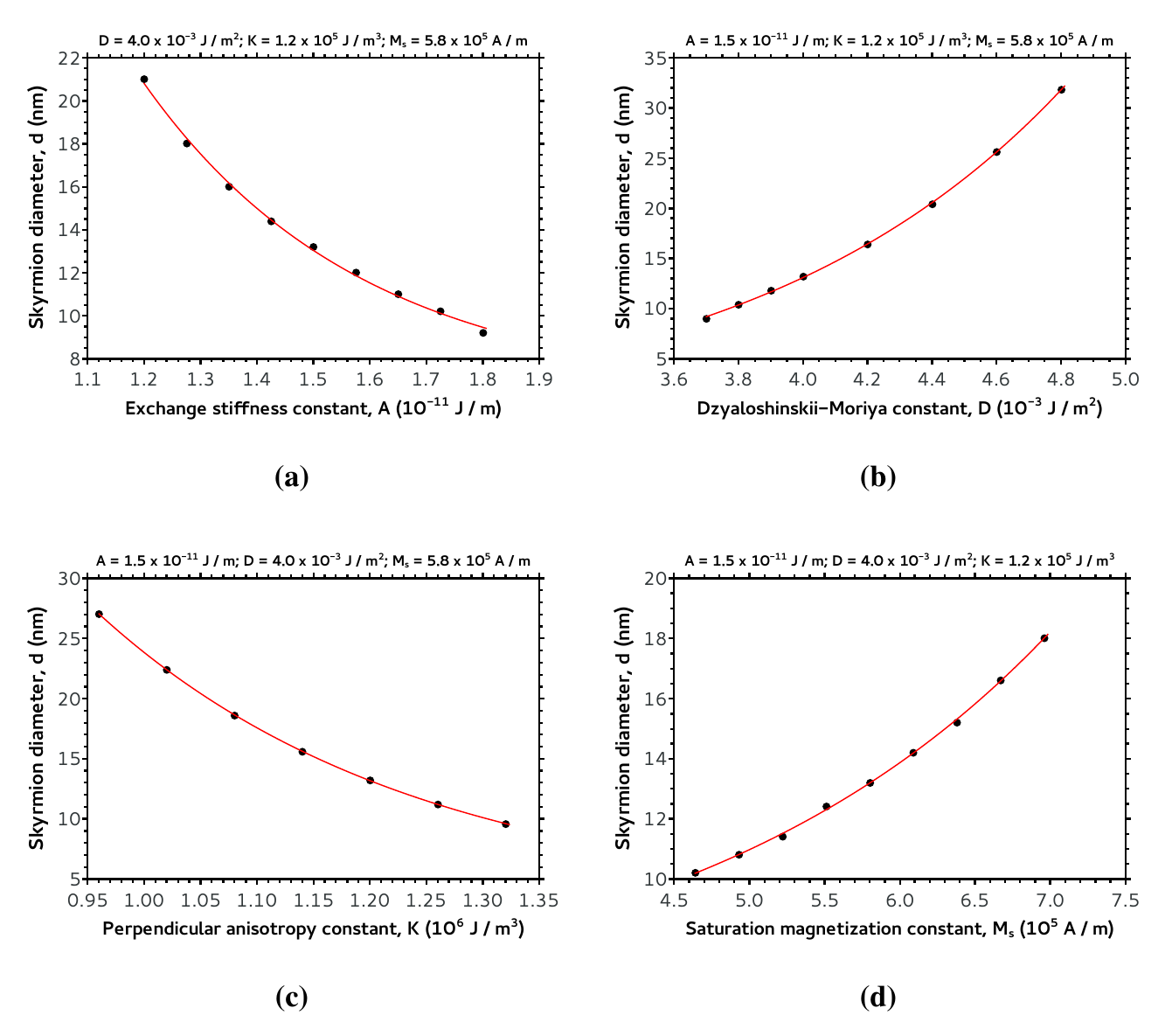}
\caption{(Color online). Skyrmion diameter as a function of the magnetic properties of the medium: (a) exchange stiffness constant, (b) Dzyaloshinskii-Moriya constant, (c) perpendicular anisotropy constant and (d) saturation magnetization constant. We fit exponential behaviors. In (a) $d(A)=d_{0,A}+d_{1,A}\:\textrm{e}^{-(A/A_{0})}$, where $d_{0,A}=6.415\:\textrm{nm}$, $d_{1,A}=320.685\:\textrm{nm}$ and $A_{0}=0.387\times 10^{-11}\:\textrm{J/m}$. In (b) $d(D)=d_{0,D}+d_{1,D}\:\textrm{e}^{(D/D_{0})}$, where $d_{0,D}=-1.646\:\textrm{nm}$, $d_{1,D}=0.245\:\textrm{nm}$ and $D_{0}=0.976\times 10^{-3}\:\textrm{J/m}^{2}$. In (c) $d(K)=d_{0,K}+d_{1,K}\:\textrm{e}^{-(K/K_{0})}$, where $d_{0,K}=2.934\:\textrm{nm}$, $d_{1,K}=736.457\:\textrm{nm}$ and $K_{0}=0.281\times 10^{6}\:\textrm{J/m}^{3}$. In (d) $d(M_{s})=d_{0,M_{s}}+d_{1,M_{s}}\:\textrm{e}^{(M_{s}/M_{s0})}$, where $d_{0,M_{s}}=5.260\:\textrm{nm}$, $d_{1,M_{s}}=0.739\:\textrm{nm}$ and $M_{s0}=2.444\times 10^{5}\:\textrm{A/m}$.}
\label{fig:skyrmions_diameter}
\end{figure}

We start our study obtaining the configuration of an isolated skyrmion in a planar nanowire. After the initial configuration was relaxed by the integration of the LLG equation we obtain skyrmions with different sizes, see Fig. (\ref{fig:skyrmions_sizes}). The skyrmion diameter is defined as the circle diameter at which the perpendicular magnetization changes of sign ($m_{z}=0$). When comparing your results with those of a previous study~\cite{NatureNanoTechnology_8_839_2013}, one can verify that they agreed with each other. If the magnetic nanostructure is large enough to host the skyrmion without it being deformed, one can note that the skyrmion diameter does not depend on the dimensions of the nanostructure. In this regime, the skyrmion diameter only depends on the parameters of the magnetic medium. The dependence of the skyrmion diameter and the magnetic properties of the medium is shown in Fig. (\ref{fig:skyrmions_diameter}). Our results for the skyrmion diameter as a function of the interaction strength of Dzyaloshinskii-Moriya and perpendicular anisotropy are consistent with those reported in Ref.~\cite{JPhysD_ApplPhys_48_115004_2015}. In our study we fit exponential behaviors for all magnetic parameters. The skyrmion diameter decreases exponentially with the strength of the parameters exchange stiffness and perpendicular anisotropy, whereas it increases exponentially with the strength of the parameters Dzyaloshinskii-Moriya and saturation magnetization. The dependence of the skyrmion size on magnetic properties of the medium can be understood by competition between exchange, Dzyaloshinskii-Moriya, perpendicular anisotropy and dipolar energies. We observe the ferromagnetic state for large values of $A$ and $K$, thus, the skyrmion diameter decreases gradually until it disappears. Therefore, the ferromagnetic alignment is favored by increasing of $A$ and $K$, so that the skyrmion diameter decreases by increasing these parameters. We observe the ferromagnetic state for small values of $D$, so that the skyrmion diameter enlarges by increasing the $D$ parameter strength. Obviously, the skyrmion shape is distorted when its diameter is larger than the width of the nanowire. To explain the dependence of the skyrmion size on the saturation magnetization, we observe that the strength of dipole-dipole interactions is directly proportional to the square of the $M_{\mbox{\tiny{S}}}$. Once we are dealing with ultrathin films, the dipolar coupling originates the shape anisotropy that work as an easy-plane anisotropy~\cite{PhysRevB_88_184422_2013}. Thus, we have the exactly opposite situation of an easy-axis anisotropy, so that the skyrmion diameter increases by increasing the $M_{\mbox{\tiny{S}}}$ parameter.     

\subsection{\label{sec:Interface}Nanotracks made of two magnetic materials}

\begin{figure}[htb!]
\centering
	\includegraphics[width=8.0cm]{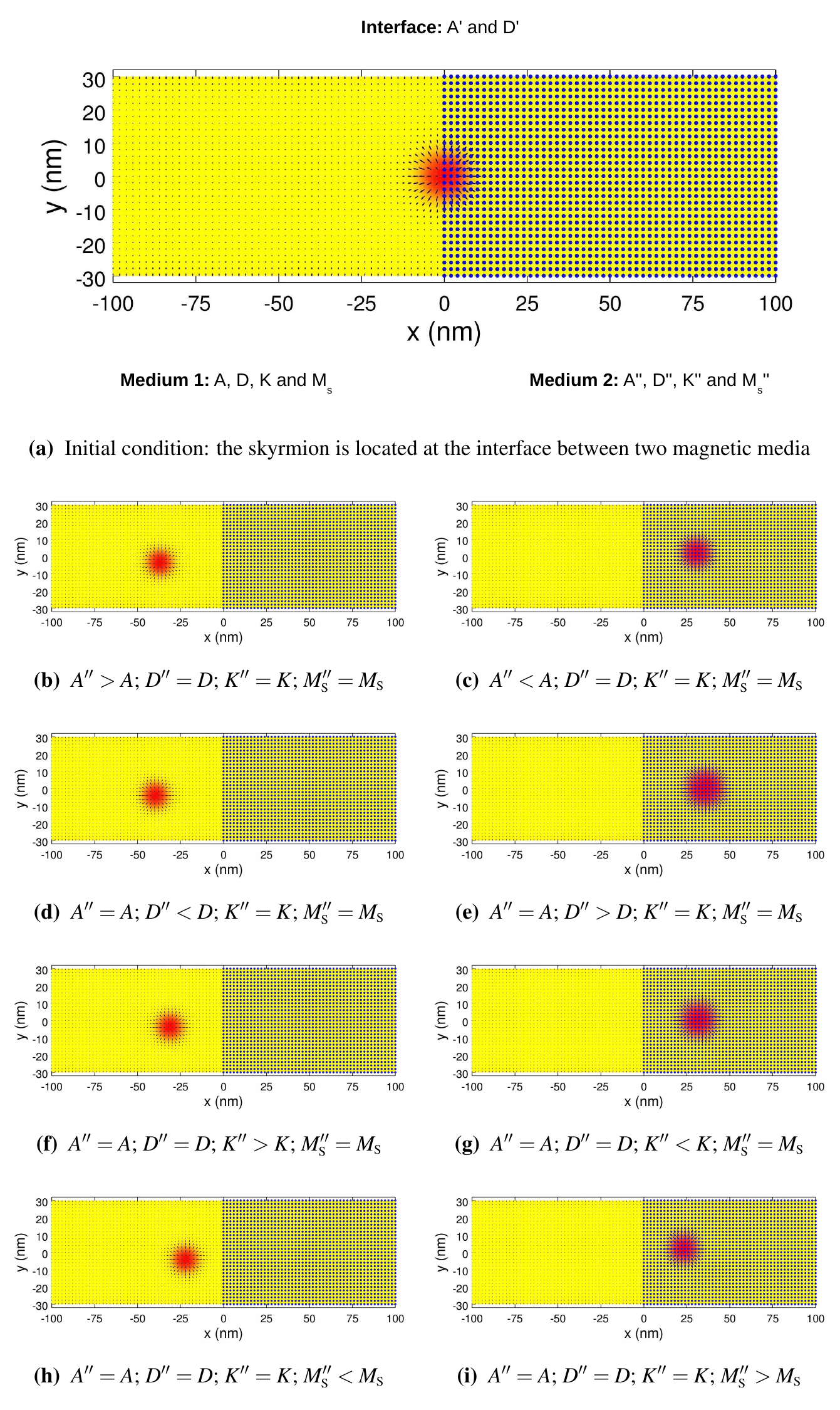}
\caption{(Color online). (a) Schematic view shows a piece the magnetic nanowire, which contains the skyrmion located at the interface between two magnetic media. At the left side of the nanowire is the medium 1 characterized by the magnetic parameters: $A, D, K, M_{\mbox{\tiny{S}}}$ and at the right side of the nanowire is the medium 2 characterized by the magnetic parameters: $A'', D'', K'', M_{\mbox{\tiny{S}}}''$. Using this configuration as initial condition, we integrate the LLG equation and obtain the final configurations shown in figures (b) to (i), which show that the skyrmion choose a medium. In figures (b) and (c) the medium 2 differs from medium 1 only in the exchange stiffness constant, $A''=1.1\: A$ and $A''= 0.9\:A$, respectively. In figures (d) and (e) the medium 2 differs from medium 1 only in the Dzyaloshinskii-Moriya constant, $D''= 0.9\:D$ and $D''=1.1\:D$, respectively. In figures (f) and (g) the medium 2 differs from medium 1 only in the perpendicular anisotropy constant, $K''=1.1\:K$ and $K''=0.9\:K$, respectively. In figures (h) and (i) the medium 2 differs from medium 1 only in the saturation magnetization constant, $M_{\mbox{\tiny{S}}}''=0.9\:M_{\mbox{\tiny{S}}}$ and $M_{\mbox{\tiny{S}}}''=1.1\:M_{\mbox{\tiny{S}}}$, respectively. Figures (b), (d), (f) and (h) show that the skyrmion moves to the medium 1, thus, it is repelled by the medium 2. On the other hand, figures (c), (e), (g) and (i) show that the skyrmion moves to the medium 2, thus, it is attracted by the medium 2.}
\label{fig:skyrmions_interface}
\end{figure}

So far we have reviewed that the balance of the magnetic interactions defines the skyrmion size. From now on, we consider a planar nanowire made of two magnetic media, as shown in Fig. \ref{fig:skyrmions_interface}(a). The magnetic properties of two media are very close, being the magnetic parameters of the medium 2 about $\pm\:10\%$ of magnetic parameters of the medium 1. The skyrmion is located exactly at the interface between two magnetic media. Using this initial condition, we numerically calculated the relaxed micromagnetic state of 60-nm-wide nanowires in zero field for different values of the magnetic parameters of the medium 2. Results of these simulations show that the skyrmion is stabilized either in the medium 1 or in the medium 2, see Fig. \ref{fig:skyrmions_interface}. From the results shown in figures (c), (e), (g) and (i), one can see that the magnetic system decreases its energy by moving the skyrmion to the medium 2, which is the region of $A$ and $K$ reduced or $D$ and $M_{\mbox{\tiny{S}}}$ increased. On the other hand, one can see that the medium 2 is avoided in figures (b), (d), (f) and (h). In these simulations, the medium 2 is the region of $A$ and $K$ increased or $D$ and $M_{\mbox{\tiny{S}}}$ reduced. To understand the two distinct behaviors, that is, the skyrmion is either attracted or repelled by the medium 2, we remember the reader of our previous results about the balance of the magnetic interactions, which defines the skyrmion size. In Fig. (\ref{fig:skyrmions_diameter}), we emphasized that the skyrmion diameter is reduced by increasing of the exchange and perpendicular anisotropy parameters, whereas it is increased by increasing of the Dzyaloshinskii-Moriya and saturation magnetization constants. Thus, we can summarize the results of the Fig. (\ref{fig:skyrmions_interface}) saying that the skyrmion prefers the magnetic region which enlarges its diameter, because it not only minimizes the system energy, but also ensures the  skyrmion survival. In other words, the system energy increases as the skyrmion diameter decreases. Once the skyrmion can disappear collapsing to the ferromagnetic state, the skyrmion avoids the magnetic region which tends to shrink its size.






\subsection{\label{sec:Defect}Nanotracks with a single magnetic defect}

\begin{figure}[htb!]
\centering
	\includegraphics[width=8.5cm]{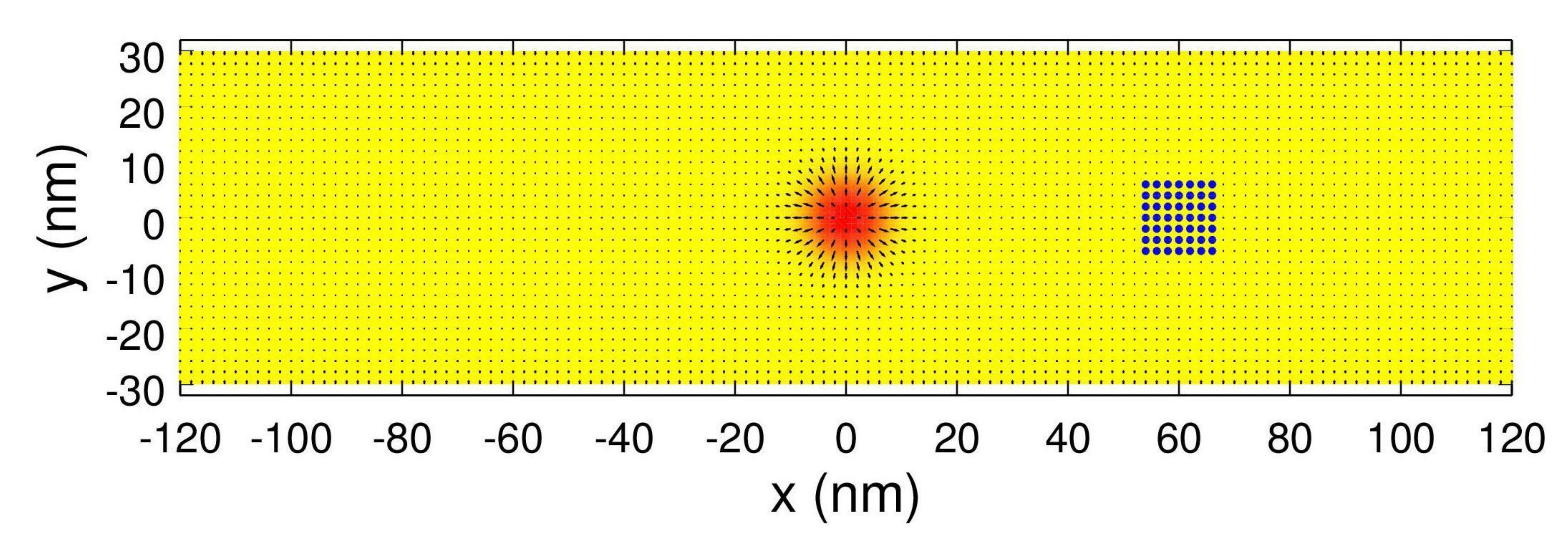}
\caption{(Color online). Schematic view shows a piece the magnetic nanowire, which contains the skyrmion and the magnetic defect. The magnetic defect consists in the local variation of the magnetic properties in the area represented by the blue spots. The majority of the wire's magnetic moments is going out of the plane of this figure except at the core of the skyrmion, where they are pointing in the opposite direction (red region). The center-to-center separation between the skyrmion and the magnetic defect is $s=60\:\textrm{nm}$. The magnetic defect has an area of 196 nm$^{2}$.}
\label{fig:Schematic}
\end{figure}
\begin{figure}[!]
\centering
	\includegraphics[width=7.5cm]{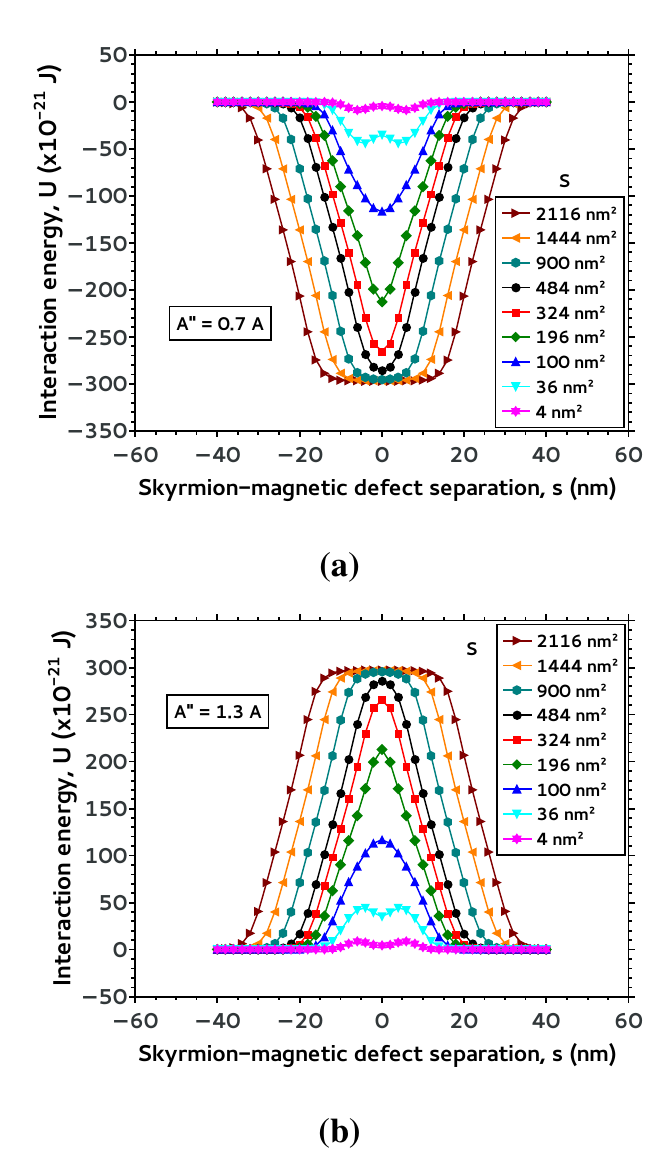}
\caption{(Color online). Local variations in the exchange stiffness constant: Type $A$ magnetic defects. Figures (a) and (b) show the interaction energy between the skyrmion and the magnetic defect as a function of the center-to-center separation. In Fig. (a) is shown a local reduction of 30\% in $A$, whereas in Fig. (b) is shown a local increase of 30\% in $A$. The behavior is shown for different areas of a type $A$ magnetic defect.}
\label{fig:A}
\end{figure}
\begin{figure}[!]
\centering
	\includegraphics[width=7.50cm]{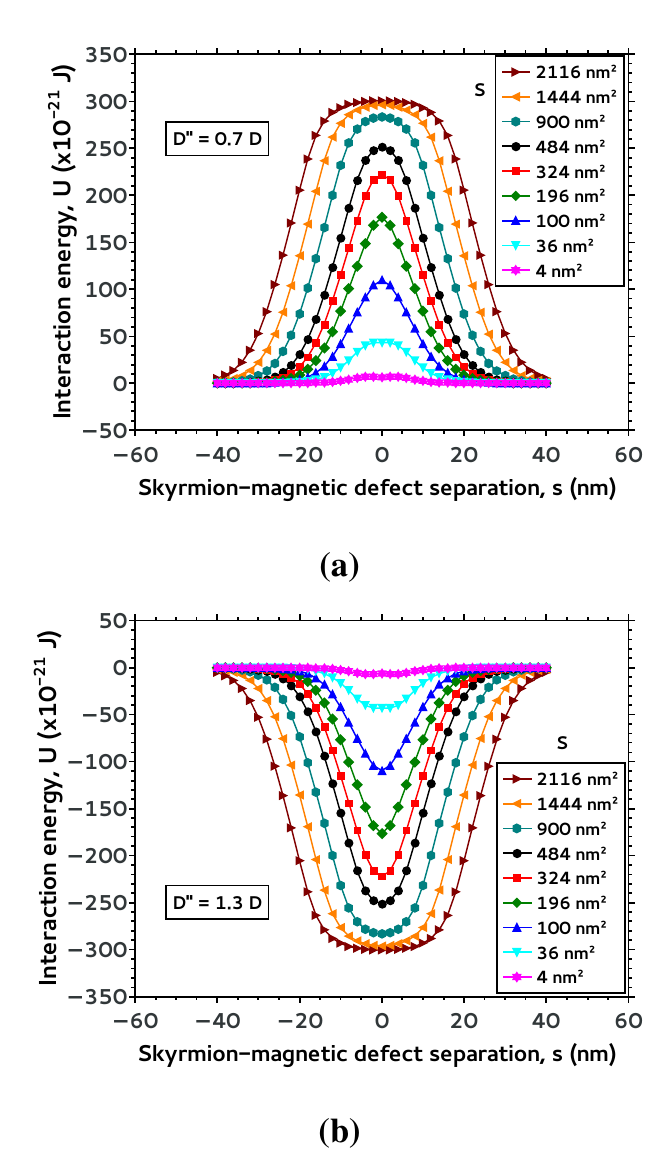}
\caption{(Color online). Local variations in the constant of the Dzyaloshinskii-Moriya interaction: Type $D$ magnetic defects. Figures (a) and (b) show the interaction energy between the skyrmion and the magnetic defect as a function of the center-to-center separation. In Fig. (a) is shown a local reduction of 30\% in $D$, whereas in Fig. (b) is shown a local increase of 30\% in $D$. The behavior is shown for different areas of a type $D$ magnetic defect.}
\label{fig:DM}
\end{figure}
\begin{figure}
\centering
	\includegraphics[width=7.50cm]{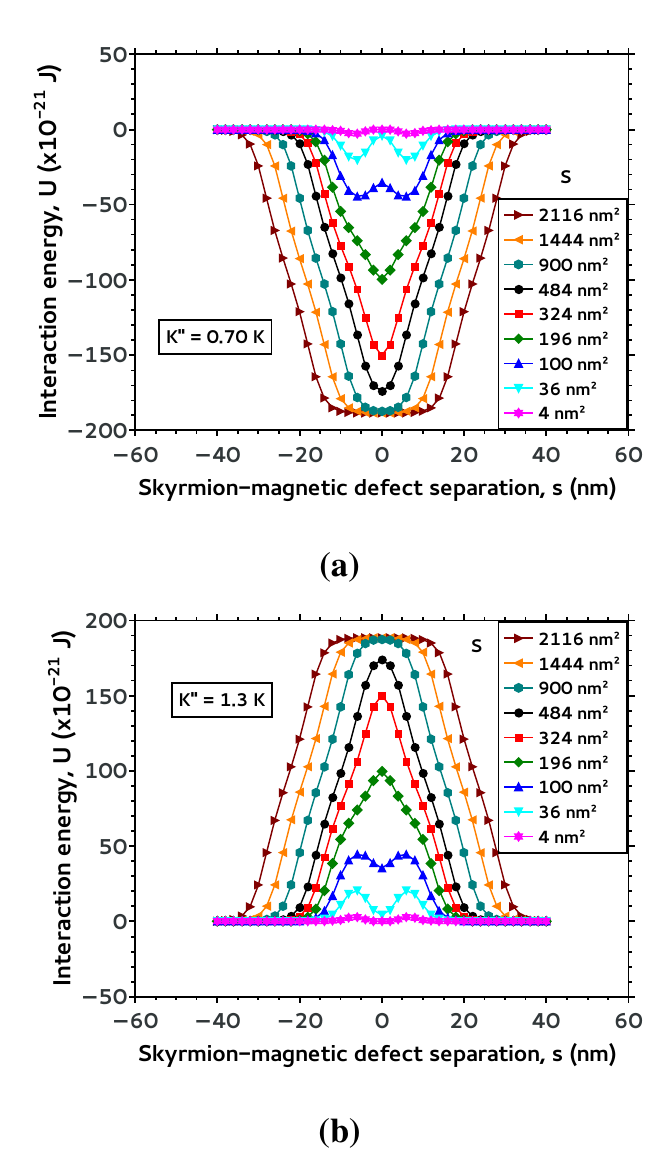}
\caption{(Color online). Local Variations in the perpendicular magnetic anisotropy: Type $K$ magnetic defects. Figures (a) and (b) show the interaction energy between the skyrmion and the magnetic defect as a function of the center-to-center separation. In Fig. (a) is shown a local reduction of 30\% in $K$, whereas in Fig. (b) is shown a local increase of 30\% in $K$. The behavior is shown for different areas of a type $K$ magnetic defect.}
\label{fig:K}
\end{figure}
\begin{figure}[htb!]
\centering
	\includegraphics[width=7.50cm]{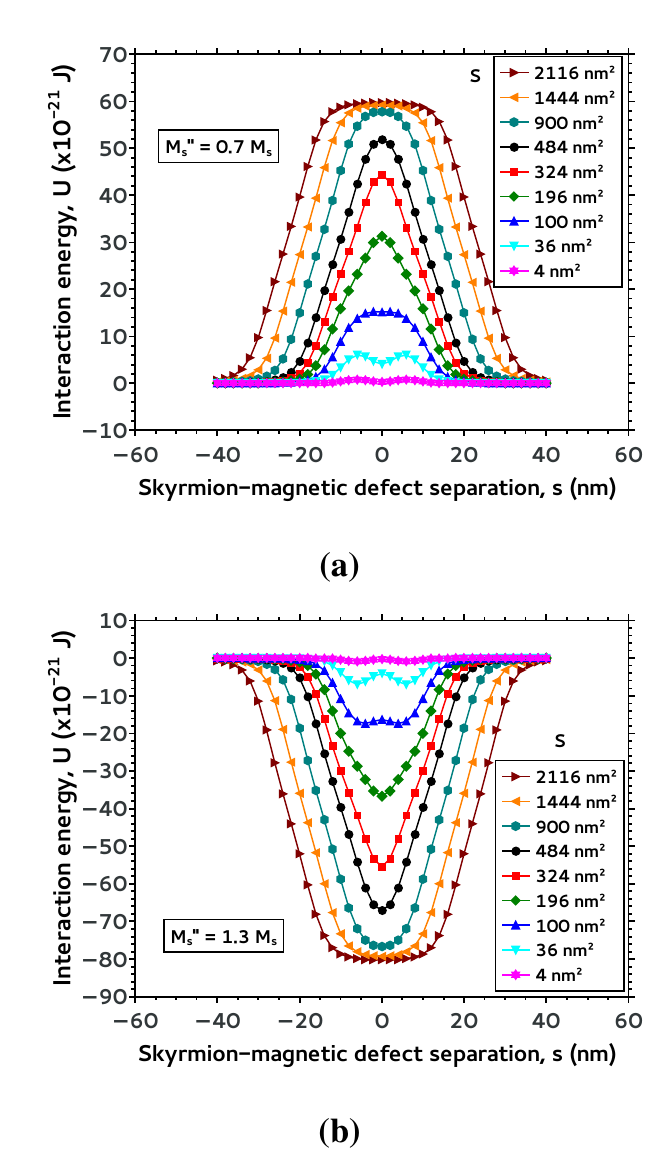}
\caption{(Color online). Local Variations in the saturation magnetization: Type $M_{\mbox{\tiny{S}}}$ magnetic defects. Figures (a) and (b) show the interaction energy between the skyrmion and the magnetic defect as a function of the center-to-center separation. In Fig. (a) is shown a local reduction of 30\% in $M_{\mbox{\tiny{S}}}$, whereas in Fig. (b) is shown a local increase of 30\% in $M_{\mbox{\tiny{S}}}$. The behavior is shown for different areas of a type $M_{\mbox{\tiny{S}}}$ magnetic defect.}
\label{fig:Ms}
\end{figure}

Now we investigate the possibility of building traps for magnetic skyrmions.  As shown in Fig. (\ref{fig:Schematic}), a trap consists in a magnetic defect incorporated into the nanostructure which hosts the skyrmion. Our study of the defect-skyrmion interaction revealed two kinds of traps. In a pinning trap, the skyrmion core moves towards the magnetic defect, indicating an effective attractive potential of interaction between the skyrmion and the magnetic defect (potential well). In a scattering trap, the skyrmion core moves away from the magnetic defect, indicating an effective repulsive potential of interaction between the skyrmion and the magnetic defect (potential barrier). Figures (\ref{fig:A}), (\ref{fig:DM}), (\ref{fig:K}), (\ref{fig:Ms}) show that both pinning and scattering traps can be individually originated by the local variation of $A, D, K, M_{\mbox{\tiny{S}}}$, respectively. From these figures, one can note that potential wells and potential barriers arise naturally as the defect size increases. Thus, magnetic defects large enough relative to the skyrmion size can be used as traps, either to pin or to scatter skyrmions. One can see that the interaction potential changes its shape for some critical size of the magnetic defect. When the defects are small, the strength of the skyrmion-defect interaction is very weak. Besides, in these cases, the interaction depends on the skyrmion-defect separation. Therefore, the skyrmion-defect aspect ratio (that is, the ratio of sizes) is a crucial parameter to design traps for skyrmions. In particular, the efficiency of the trap is compromised if the defect size is smaller than the skyrmion size.

%
%
%
%
%
%
%
%
%




%



\newpage

\begin{figure}[htb!]
\centering
	\includegraphics[width=8.2cm]{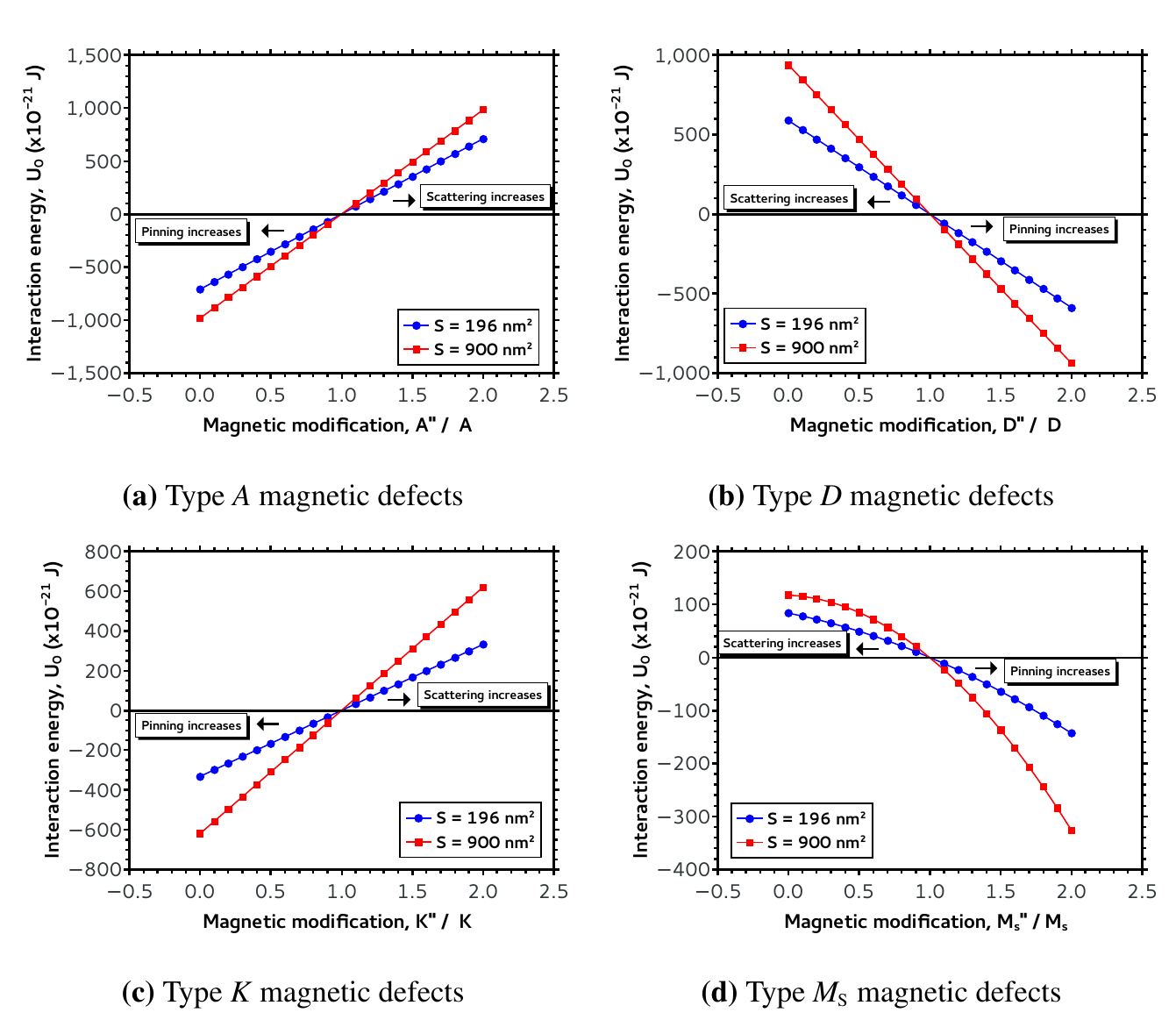}
\caption{(Color online). Interaction energy $U_{0}$ as a function of a local variation in the magnetic properties. The behavior is shown for different types of magnetic defects.
Linear and quadratic dependencies of the skyrmion-defect interaction with the located magnetic modifications do not surprise, once the same dependencies are presents in the strength of the magnetic interactions, it can be confirmed in Eqs. (\ref{eq:A}), (\ref{eq:D}), (\ref{eq:K}) and (\ref{eq:Ms}).}
\label{fig:U0_X}
\end{figure}

\begin{figure}[htb!]
\centering
	\includegraphics[width=8.2cm]{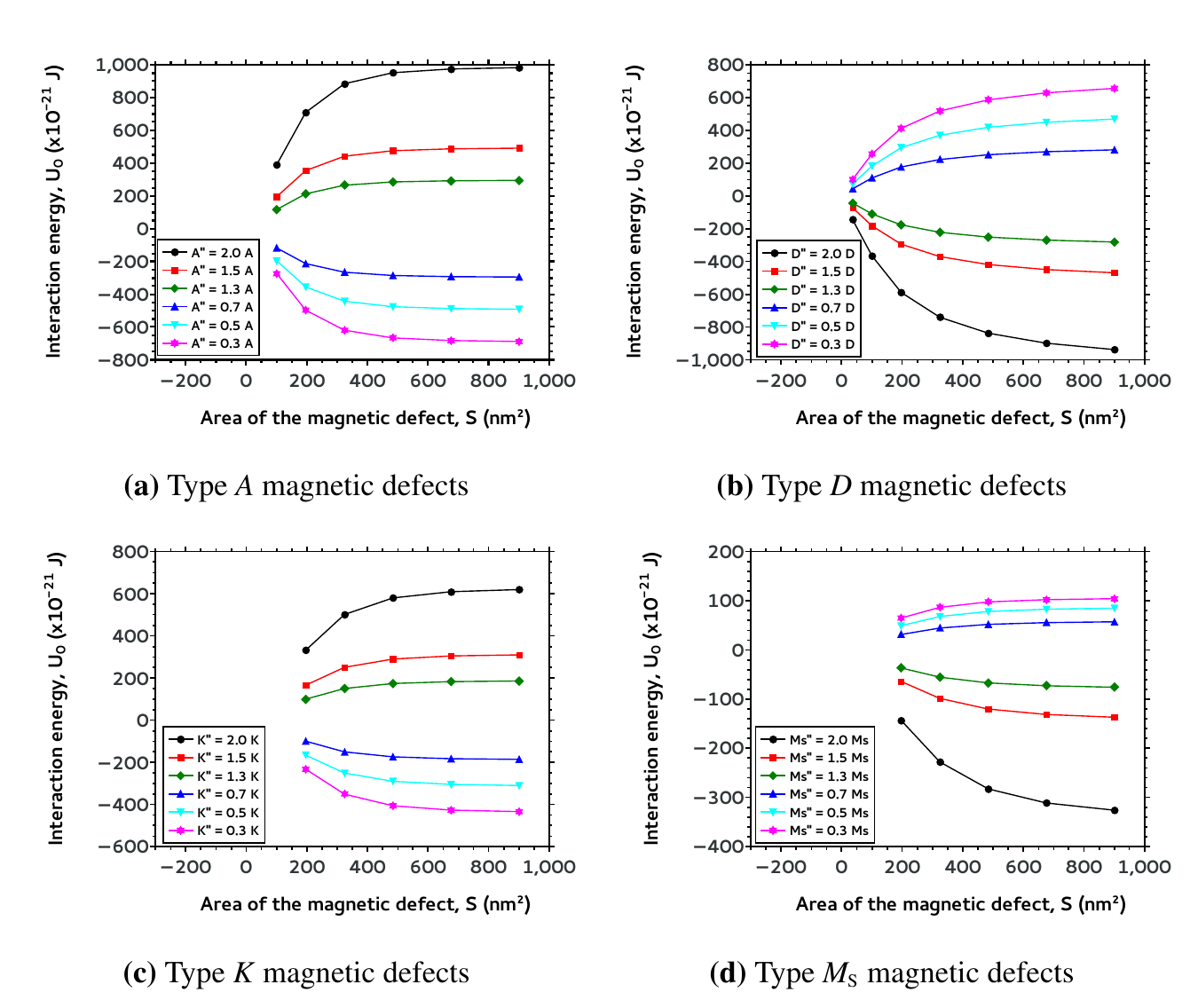}
\caption{(Color online). Interaction energy $U_{0}=U(s=0)$ as a function of the defect area $S$. The behavior is shown for different types of magnetic defects. 
Exponential behaviors have been fitted for all types of defects. The missing points in the curves for defects of small areas represent the situation in which the graph of the interaction potential does not correspond neither a potential well nor a potential barrier. As has been mentioned previously, the traps do not work for magnetic defect of small areas.}
\label{fig:U0_area}
\end{figure}

\begin{figure}[htb!]
\centering
	\includegraphics[width=8.2cm]{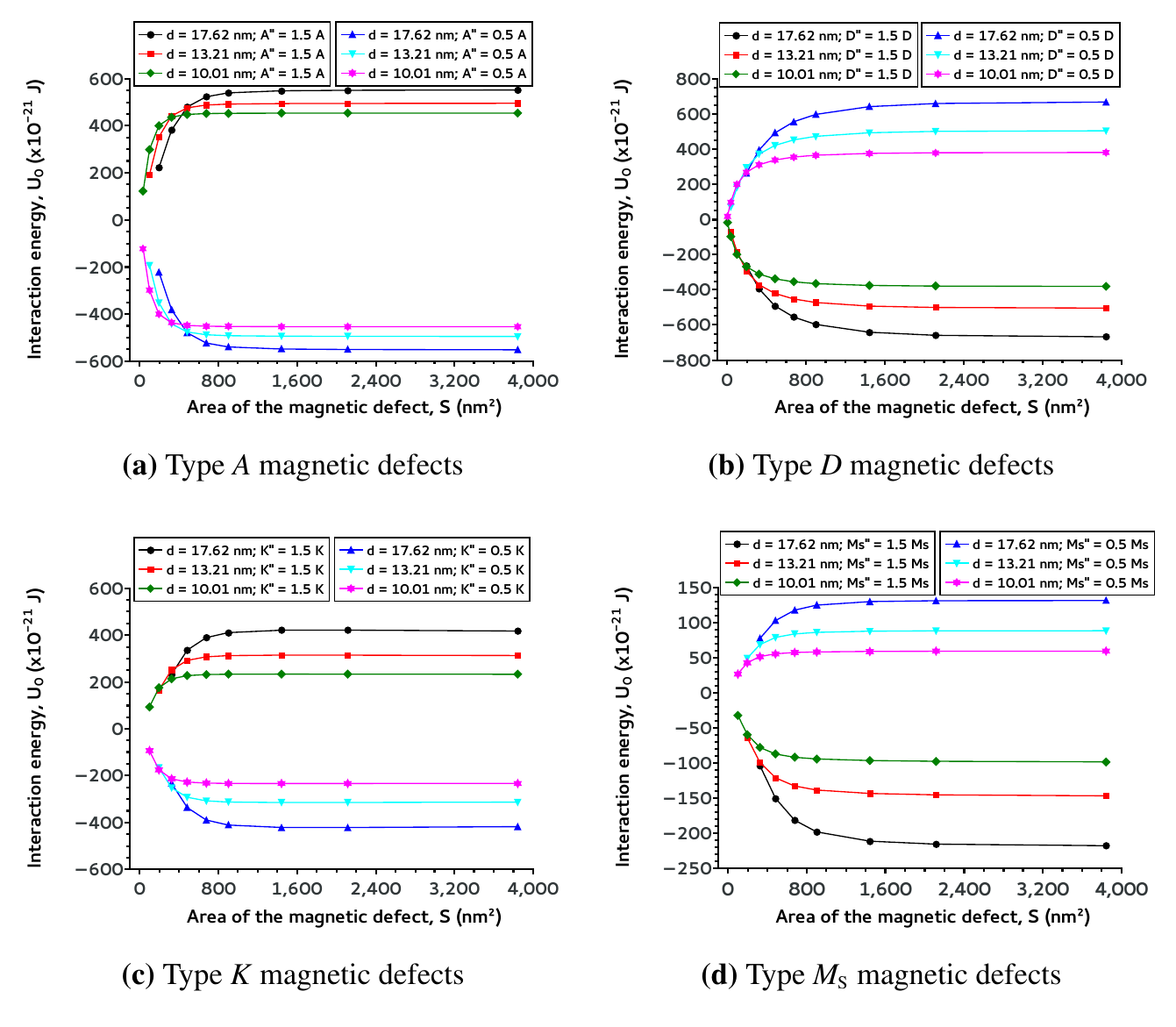}
\caption{(Color online). Interaction energy $U_{0}$ as a function of the defect area $S$ and the skyrmion diameter $d$. The behavior is shown for different types of magnetic defects. 
The missing points in the curves for defects of small areas represent the situation in which the graph of the interaction potential does not correspond neither a potential well nor a potential barrier. As has been mentioned previously, the traps do not work for magnetic defect of small areas.}
\label{fig:U0_diameter}
\end{figure}

In order to systematically understand the skyrmion-defect interaction, we define $U_{0}=U(s=0)$ which measures the strength of the interaction, that is, the maximum value or the minimum value of $U$.  If $U_{0}>0$, then the maximum value $U_{0}$ represents the height of the potential barrier, whereas if $U_{0}<0$, then the minimum value $\vert U_{0}\vert$ represents the depth of the potential well. As shown in Figs. (\ref{fig:U0_X}), (\ref{fig:U0_area}) and (\ref{fig:U0_diameter}), the strength of the skyrmion-defect interaction depends on the local variations of the magnetic property as well as on the skyrmion-defect aspect ratio. In particular,  from figures (\ref{fig:U0_area}) and (\ref{fig:U0_diameter}), one can see that the strength of the interaction increases not only by increasing of the area of the magnetic defect, but also by increasing of the skyrmion size. Thus, the strength of the skyrmion-defect interaction can be tuned by the modification of the magnetic properties within a region with suitable size. It is notably that the efficiency of the trap is improved as the skyrmion size becomes much smaller than the defect size.





%

In Table (\ref{tab:traps}) we summarize four distinct ways to build traps for pinning and scattering magnetic skyrmions. Although the discussions have been held with the N\'{e}el skyrmion, we would like to emphasize that the qualitative results of the Table (\ref{tab:traps}) remain unchanged for Bloch skyrmions; it has been checked through micromagnetic simulations. Moreover, we have verified that the results remain essentially the same for both chirality of the skyrmion ($D>0$ and $D<0)$. 
 
\begin{table}[htb!]
\caption{Two types of traps for magnetic skyrmions can be originated in located variations of the magnetic properties when tuning either a local increase ($X''\:>\:X$) or a local reduction ($X''\:<\:X$), where $X$ can be $A, D, K \:\textrm{or}\: M_{\mbox{\tiny{S}}}$. A pinning trap corresponds to a potential well for the skyrmion, whereas a scattering trap corresponds to a potential barrier.}
%
%
%
%
\centering
\vspace{0.5cm}
\begin{tabular}{|c|>{\centering\arraybackslash}m{0.7in}|>{\centering\arraybackslash}m{0.7in}|}

%
\hline 
%
\textbf{Magnetic Property}          &  \textbf{Pinning Trap} &  \textbf{Scattering Trap} \\
\hline 
\small{exchange stiffness}               &          $A'' <  A$   &        $A'' >  A$  \\
\hline 
\small{Dzyaloshinskii-Moriya constant}   &           $D''\:>\:D $ &        $D''\: <\:D$ \\
\hline 
\small{perpendicular anisotropy}         &           $K'' \: <\: K$     &  $K''\:> \: K$  \\
\hline 
\small{saturation magnetization}   &  $M_{\mbox{\tiny{S}}}''\:>\:M_{\mbox{\tiny{S}}} $  &  $M_{\mbox{\tiny{S}}}'' \: < M_{\mbox{\tiny{S}}}$  \\
\hline 
\end{tabular}
\label{tab:traps}
\end{table}

\begin{figure}[htb!]

\centering
	\includegraphics[width=8.2cm]{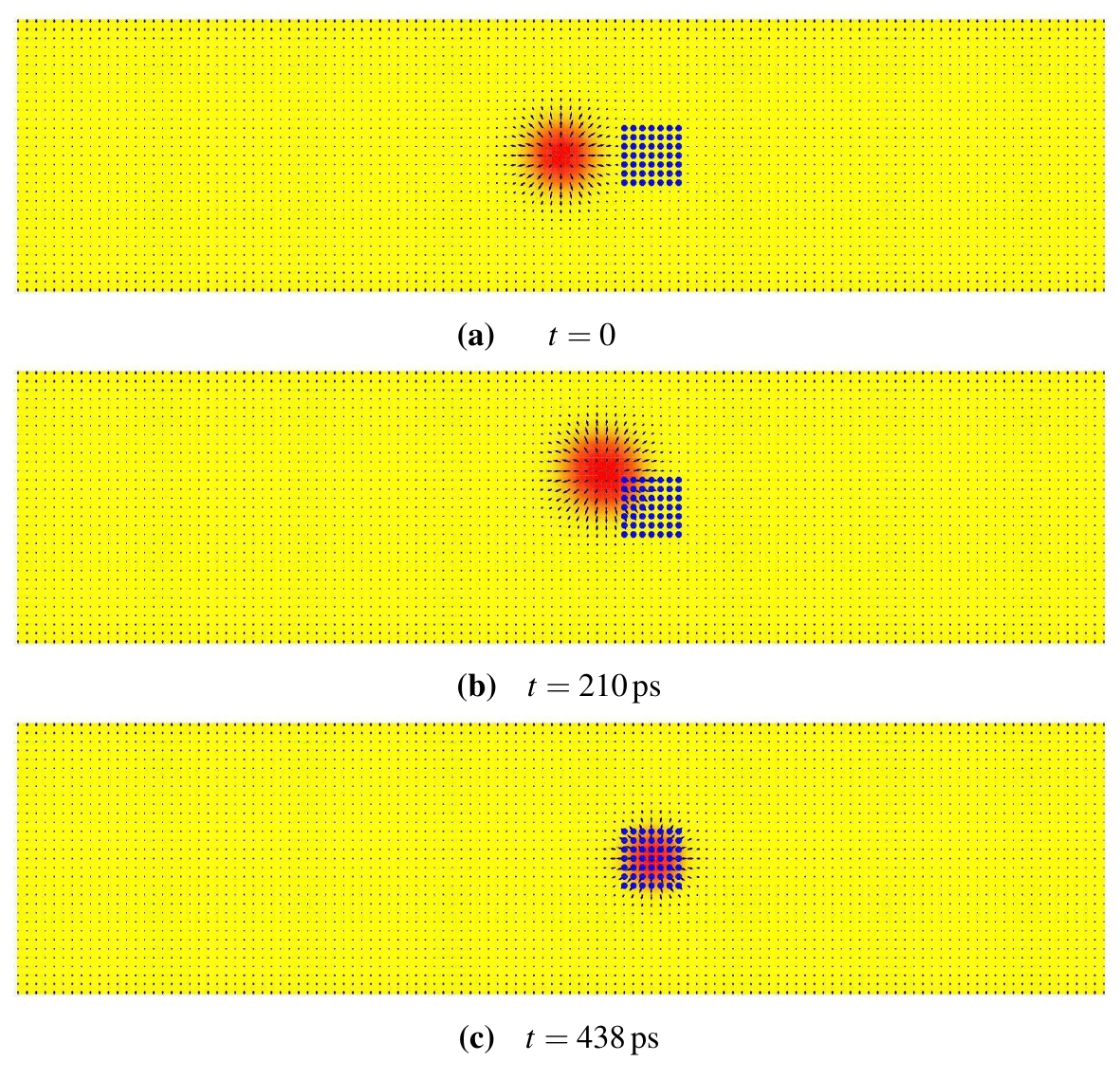}
	\caption{(Color online). Effect of a pinning trap. Successive snapshots show the skyrmion being attracted by a trap, which is characterized by a local reduction in the exchange stiffness constant $A''=0.7\:A$ (a local reduction of 30 \% in A) into an area of 196 nm$^{\:2}$. (a) Initial configuration at t = 0, where the center-to-center separation between the skyrmion and the magnetic defect is 20 nm. (b) Configuration after 210 ps. (c) Configuration after 438 ps, where the skyrmion is captured by the pinning trap.}
   \label{fig:Pinning_Trap}
\end{figure}
\begin{figure}[htb!]

\centering
	\includegraphics[width=8.2cm]{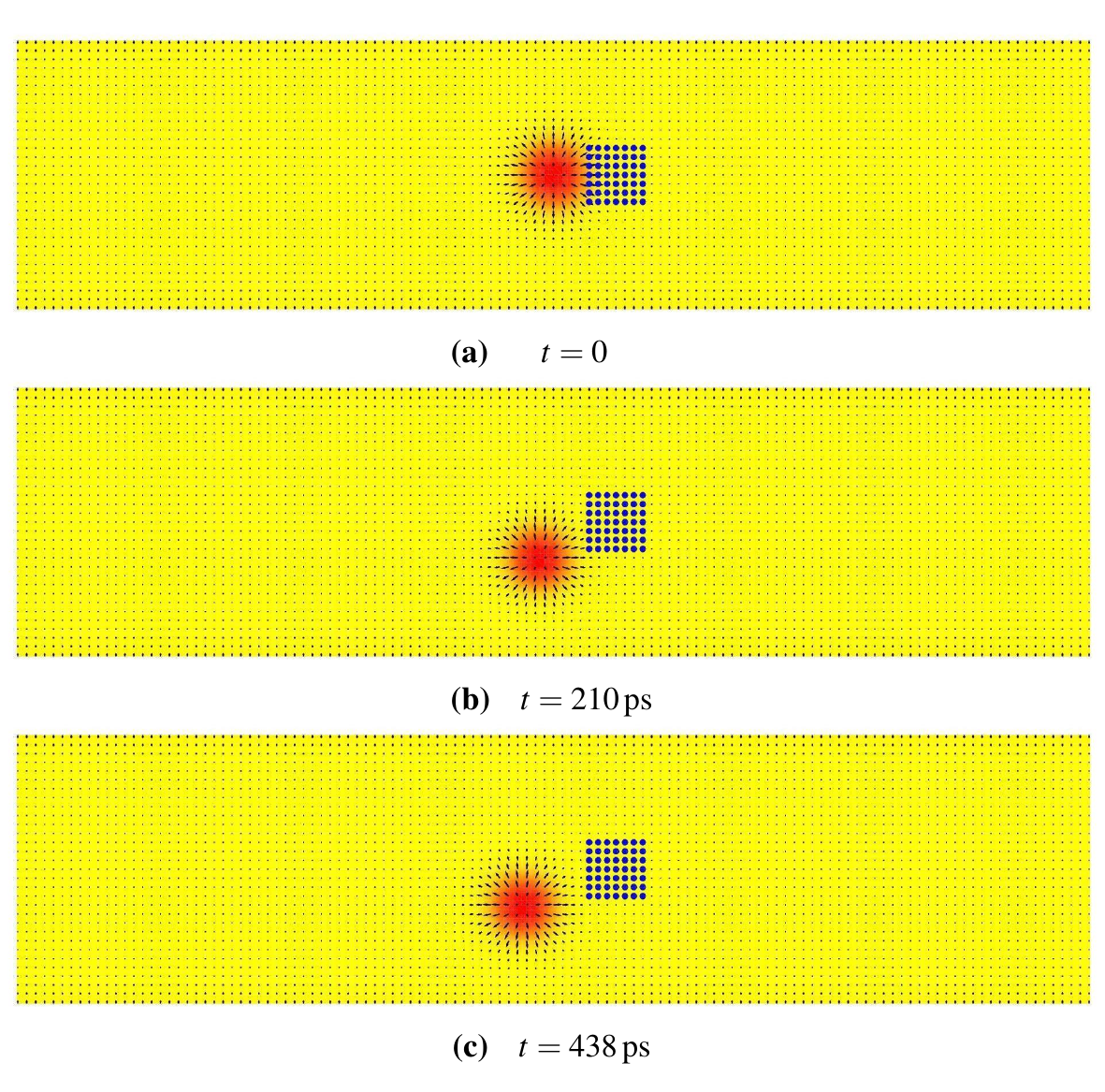}
	\caption{(Color online). Effect of a scattering trap. Successive snapshots show the skyrmion being repelled by a trap, which is characterized by a local increase in the exchange stiffness constant $A''=1.3\:A$ (a local increase of 30 \% in A) into an area of 196 nm$^{\:2}$. (a) Initial configuration at t = 0, where the center-to-center separation between the skyrmion and the magnetic defect is 14 nm. (b) Configuration after 210 ps. (c) Configuration after 438 ps.}
   \label{fig:Scattering_Trap}
\end{figure}

Now we verify the predictions of the figure (\ref{fig:A}) through micromagnetic simulations, see figures (\ref{fig:Pinning_Trap}) and (\ref{fig:Scattering_Trap}). By considering a magnetic defect with a suitable size, we show that the skyrmion can be either attracted by a reduced $A$ region or repelled by an increased $A$ region. Although the results presented in figures (\ref{fig:Pinning_Trap}) and (\ref{fig:Scattering_Trap}) are for the case of Type $A$ magnetic defects, we have observed similar results (both pinning and scattering traps) for other types of magnetic defects which were approached in this work, see Table (\ref{tab:traps}).

\newpage

It is also interesting to discuss the influence of the defect size on the magnitude of the skyrmion-defect interaction. As shown in figures (\ref{fig:A}), (\ref{fig:DM}), (\ref{fig:K}) and (\ref{fig:Ms}), the interaction potential can present a double-well potential (or double-barrier potential) instead of a single potential well (or a single potential barrier). Particularly, this situation occurs whenever the defect is smaller than the skyrmion and the interaction between them is extremely weak. Besides, micromagnetic simulations show that the skyrmion pinning due to a double-well potential generates a pinning center which does not coincide with the defect center. 

From the technological point of view, the reader can wonder whether magnetic defects considered in this work would work as traps for skyrmions in a nanotrack at room temperature? By analyzing our results for skyrmion-defect interaction energy, one can see that the heights of the potential barriers as well as the depths of the potential wells depend strongly on the type of defect being considered, on the strength of the modified magnetic property, and on the defect size (skyrmion-defect aspect ratio). Thus, characteristics of trap can be experimentally controlled. For instance, the skyrmion-defect interaction strength can be tuned in one or more order of magnitude larger than the thermal energy, that is, $E_{\:\textrm{thermal}} \sim 10^{\:-21} [\textrm{J}] \sim k_{B}\:\:T_{\textrm{room}}$.


It is worth mentioning that the region with modified magnetic properties can be designed to present a single or a combination of tuned magnetic parameters. The balance of the magnetic defect parameters will be responsible for adjusting a proper trap, either to capture or to blockade the skyrmions. Thus, the kind of the trap depends on the suitable choice which material parameters will be modified within the selected area. Maybe, the affected region by ion beam irradiation contains more than one modified magnetic property. However, the magnitude of the affected magnetic parameters will not be the same; they can differ even in the order of magnitude. From Figs. (\ref{fig:A}), (\ref{fig:DM}), (\ref{fig:K}) and (\ref{fig:Ms}), one can see that the individual contribution of the magnetic parameter $A, D, \textrm{and}\: K$ are of the same magnitude order ($\sim 10^{\:2}\:\: E_{\:\textrm{thermal}}$), whereas the  $M_{\mbox{\tiny{S}}}$ parameter modification results in the weakest skyrmion-defect interaction ($\sim 10\:\: E_{\:\textrm{thermal}}$). Our results could guide the design of experimental works that intend to investigate the realistic incorporation of such magnetic defects into nanomagnets.

%
\section{\label{sec:Conclusion}Conclusion}
Generally, the interaction between a skyrmion and a magnetic defect is difficult of being observed because of two very simple reasons:
(1) Skyrmions are topologically protected entities. Thus, these quasiparticles have a kind of shield, which hinders the skyrmion interaction with inhomogeneities of the magnetic medium. Ultimately, due to the topological protection of the skyrmion, the skyrmion-defect interaction is weak by nature. 
(2) Up to now, most of the theoretical and computational works has considered magnetic defects with very small sizes. Besides masking the skyrmion-defect interaction, the incorporation of small defects in nanomagnets can be a challenge for current experimental techniques.

In this work, we show that the skyrmion-defect interaction becomes evident as the defect size increases. Both attractive and repulsive interactions have been observed whenever the defect size was larger than the skyrmion size. Therefore, traps for magnetic skyrmions can be created through magnetic defects with sizes large enough. Indeed, we conclude that the skyrmion-defect aspect ratio is a crucial parameter to build traps for skyrmions, which can be captured or blocked around a magnetic defect intentionally incorporated into the nanomagnet.
%
%
Magnetic defects have been modeled as located variations on the material parameters, such as $A, D, K, M_{\mbox{\tiny{S}}}$. Through micromagnetic simulations we show that any of these material parameters can be used to create both pinning and scattering traps by tuning either a local increase or a local decrease of a given magnetic property. In order to provide a background for experimental studies, we have varied the parameters of the magnetic defects. In contrast to previous studies on the skyrmion-defect interaction, we consider the effect of the full dipolar coupling. All the possibilities for a skyrmion trap were investigated and they are summarized in Table (\ref{tab:traps}). Furthermore, we understand the basic physics behind the mechanisms for pinning and for scattering magnetic skyrmions, that is, skyrmions move towards the  magnetic region which tends to maximize its diameter. It enables the magnetic system to minimize its energy. Thus, we are able to explain why skyrmions are either attracted or repelled by a region with modified magnetic properties. 

We believe that it is possible to engineer magnetic defects working as skyrmion traps, and we are hoping for this paper encourages future experimental works to investigate the skyrmion-defect interaction as well as to verify our predictions. As we emphasized, the magnetic defect size must be large enough relative to the skyrmion size, somehow, it facilitates the realistic incorporation of such magnetic defects into nanomagnets.

Understanding and controlling the static properties and dynamics of skyrmions in magnetic nanostructures is of utmost significance for the development and realization of future spintronic devices. Although the results presented here are for a very simple distribution of magnetic defects into a ferromagnetic nanotrack, we believe their consequences can be planned and extended. For instance, a nanotrack containing a proper distribution of magnetic defects can be designed to solve the issue of transporting skyrmions in a usual ferromagnetic nanowire. 




%

\section*{\label{sec:Acknowledgments}Acknowledgments}
This work was partially supported by CAPES, CNPq, FAPEMIG and FINEP (Brazilian Agencies). Numerical simulations were done at the Laborat\'{o}rio de Simula\c{c}\~{a}o Computacional do Departamento de F\'{\i}sica da UFJF.
%



\begin{thebibliography}{100}



%
\bibitem{Proc_R_Soc_London_Ser_A_260_127_1961}{T. H. R. Skyrme, Proc. R. Soc. London Ser. A {\bf 260}, 127-138 (1961).}
%
\bibitem{Nucl_Phys_31_556_1962}{T. H. R. Skyrme, Nucl. Phys. {\bf 31}, 556-569 (1962).}
%
%
%
%
\bibitem{Rev_Mod_Phys_61_385_1989}{D. C. Wright and N. D. Mermin, Rev. Mod. Phys. {\bf 61}, 385-432 (1989).}
%
\bibitem{Nat_Commun_2_246_2011}{Jun-ichi Fukuda and Slobodan Zumer, Nat. Commun. {\bf 2}, 246 (2011).}
%
%
\bibitem{PhysRevLett_81_742_1998}{Tin-Lun Ho, Phys. Rev. Lett. {\bf 81}, 742-745 (1998).}
%
\bibitem{Nat_London_411_918_2001}{U. Al Khawaja and H. Stoof, Nature (London) {\bf 411}, 918-920 (2001).}
%
\bibitem{New_J_Phys_17_069501_2015}{J.-y. Choi, W. J. Kwon, M. Lee, H. Jeong, K. An, and Y.-il Shin, New J. Phys. {\bf 17}, 069501 (2015).}
%
%
%
\bibitem{PhysRevLett_119_167001_2017}{A. A. Zyuzin, J. Garaud, and E. Babaev, Phys. Rev. Lett. {\bf 119}, 167001 (2017).}
%
\bibitem{JPhys_CondensMatter_30_295601_2018}{C. Morice, D. Chakraborty, X. Montiel, and C. P\'{e}pin, J. Phys.: Condens. Matter. {\bf 30}, 295601 (2018).}

%
%
%


%
\bibitem{Belavin_Polyakov_1975}{A. A. Belavin and A. M. Polyakov, JETP Lett. {\bf 22}, 245 (1975).}
%
\bibitem{JMagnMagnMater_138_255_1994}{A. Bogdanov and A. Hubert, J. Magn. Magn. Mater. {\bf 138}, 255-269 (1994).}
%
\bibitem{JMagnMagnMater_305_413_2006}{A. S. Kirakosyan and V. L. Pokrovsky, J. Magn. Magn. Mater. {\bf 305}, 413-422 (2006).}

\bibitem{Nature_442_797_2006}{U. K. R\"{o}$\beta$ler, A. N. Bogdanov and C. Pfleiderer, Nature {\bf 442}, 797-801 (2006).}
%
\bibitem{NatureNanoTechnology_8_839_2013}{J. Sampaio, V. Cros, S. Rohart, A. Thiaville, and A. Fert, Nat. Nanotechnol. {\bf 8}, 839-844 (2013).}
%
\bibitem{NatureNanoTechnology_8_899_2013}{N. Nagaosa and Y. Tokura, Nat. Nanotechnol. {\bf 8}, 899-911 (2013).}



\bibitem{Nat_Mater_6_813_2007}{C. Chappert, A. Fert and F. N. Van Dau, Nat. Mater. {\bf 6}, 813-823 (2007).}
%
\bibitem{Nat_Rev_Mater_2_17031_2017}{A Fert, N. Reyren and V. Cros, Nat. Rev. Mater. {\bf 2}, 17031 (2017).}



%
%
\bibitem{JPhysD_ApplPhys_44_392001_2011}{N. S. Kiselev, A. N. Bogdanov, R. Sch\"{a}fer, and U. K. R\"{o}$\beta$ler, J. Phys. D: Appl. Phys. {\bf 44}, 392001 (2011).}
%
%
\bibitem{NatureNanoTechnology_8_152_2013}{A. Fert, V. Cros and J. Sampaio, Nat. Nanotechnol. {\bf 8}, 152-156 (2013).}

%


%
\bibitem{Science_323_915_2009}{S. M\"{u}hlbauer, B. Binz, F. Jonietz, C. Pfleiderer, A. Rosch, A. Neubauer, R. Georgii, and P. B\"{o}ni, Science {\bf 323}, 915-919 (2009).}
%
\bibitem{PhysRevB_81_041203_2010}{W. M\"{u}nzer, A. Neubauer, T. Adams, S. M\"{u}hlbauer, C. Franz, F. Jonietz, R. Georgii, P. B\"{o}ni, B. Pedersen, M. Schmidt, A. Rosch, and C. Pfleiderer, Phys. Rev. B {\bf 81}, 041203(R) (2010).}
%
\bibitem{Nature_465_901_2010}{X. Z. Yu, Y. Onose, N. Kanazawa, J. H. Park, J. H. Han, Y. Matsui, N. Nagaosa, and Y. Tokura, Nature {\bf 465}, 901-904 (2010).}
%
\bibitem{NaturePhysics_7_713_2011}{S. Heinze, K. von Bergmann, M. Menzel, J. Brede, A. Kubetzka, R. Wiesendanger, G. Bihlmayer, and S. Bl\"{u}ge, Nature Phys. {\bf 7}, 713-718 (2011).}
%
\bibitem{Nat_Mater_10_106_2011}{X. Z. Yu, N. Kanazawa, Y. Onose, K. Kimoto, W. Z. Zhang, S. Ishiwata, Y. Matsui, and Y. Tokura, Nat. Mater. {\bf 10}, 106-109 (2011).}

%
\bibitem{Science_336_198_2012}{S. Seki, X. Z. Yu, S. Ishiwata, and Y. Tokura Science {\bf 336}, 198-201 (2012).}
%
\bibitem{Science_341_636_2013}{N. Romming, C. Hanneken, M. Menzel, J. E. Bickel, B. Wolter, K. von Bergmann, A. Kubetzka, and R. Wiesendanger, Science {\bf 341}, 636-639 (2013).}



%

%
\bibitem{Science_349_283_2015}{W. Jiang, P. Upadhyaya, W. Zhang, G. Yu, M. B. Jungfleisch, F. Y. Fradin, J. E. Pearson, Y. Tserkovnyak, K. L. Wang, O. Heinonen, S. G. E. te Velthuis, and A. Hoffmann, Science {\bf 349}, 283-286 (2015).}
%
\bibitem{ApplPhysLett_106_242404_2015}{G. Chen, A. Mascaraque, A. T. N'Diaye, and A. K. Schmid, Appl. Phys. Lett. {\bf 106}, 242404 (2015).}
%
\bibitem{Nat_Mater_15_501_2016}{S. Woo, K. Litzius, B. Kr\"{u}ger, M-Y. Im, L. Caretta, K. Richter, M. Mann, A.Krone, R. M. Reeve, M. Weigand, P. Agrawal, I. Lemesh, M-A. Mawass, P. Fischer, M. Kl\"{a}ui, and G. S. D. Beach, Nat. Mater. {\bf 15}, 501-506 (2016).}
\bibitem{NatureNanoTechnology_11_444_2016}{C. M.-Luchaire, C. Moutafis, N. Reyren, J. Sampaio, C. A. F. Vaz, N. Van Horne, K. Bouzehouane, K. Garcia, C. Deranlot, P. Warnicke, P. Wohlh\"{u}ter, J.-M. George, M. Weigand, J. Raabe, V. Cros, and A. Fert, Nat. Nanotechnol. {\bf 11}, 444-448 (2016).}
%
\bibitem{NatureNanoTechnology_11_449_2016}{O. Boulle, J. Vogel, H. Yang, S. Pizzini, D. S. Chaves, A. Locatelli, T. O. Mentes, A. Sala, L. D. Buda-Prejbeanu, O. Klein, M. Belmeguenai, Y. Roussign\'{e}, A. Stashkevich, S. Mourad Ch\'{e}rif, L. Aballe, M. Foerster, M. Chshiev, S. Auffret, I. M. Miron, and G. Gaudin, Nat. Nanotechnol. {\bf 11}, 449-454 (2016).}
%
\bibitem{ApplPhysLett_111_202403_2017}{M. He, L. Peng, Z. Zhu, G. Li, J. Cai, J. Li, H. Wei, L. Gu, S. Wang, T. Zhao, B. Shen, and Y. Zhang, Appl. Phys. Lett. {\bf 111}, 202403 (2017).}
%
\bibitem{ApplPhysLett_112_132405_2018}{S. Zhang, J. Zhang, Q. Zhang, C. Barton, V. Neu, Y. Zhao, Z. Hou, Y. Wen, C. Gong, O. Kazakova, W. Wang, Y. Peng, D. A. Garanin, E. M. Chudnovsky, and X. Zhang, Appl. Phys. Lett. {\bf 112}, 132405 (2018).}


\bibitem{Dzyaloshinskii_J_Phys_Chem_Solids_4_241_1958}{I. Dzyaloshinsky, J. Phys. Chem. Solids {\bf 4}, 241-255 (1958).}
%
\bibitem{Moriya_PhysRev_120_91_1960}{T. Moriya, Phys. Rev. {\bf 120}, 91-98 (1960).}
%
\bibitem{JMagnMagnMater_182_341_1998}{A. Cr\'{e}pieux and C. Lacroix, J. Magn. Magn. Mater. {\bf 182}, 341-349 (1998).}
%
\bibitem{PhysRevLett_87_037203_2001}{A. N. Bogdanov and U. K. R\"{o}$\beta$ler, Phys. Rev. Lett. {\bf 87}, 037203 (2001).}
%
\bibitem{JPhys_CondensMatter_24_086001_2012}{C. D. Hu, J. Phys.: Condens. Matter. {\bf 24}, 086001 (2012).}
%
%

%
%






\bibitem{PhysRevLett_105_197202_2010}{Motohiko Ezawa, Phys. Rev. Lett. {\bf 105}, 197202 (2010).}
%
\bibitem{PhysRevB_88_054403_2013}{Y. Y. Dai, H. Wang, P. Tao, T. Yang, W. J. Ren, and Z. D. Zhang, Phys. Rev. B {\bf 88}, 054403 (2013).}







%

\bibitem{NatureNanoTechnology_8_723_2013}{K. Shibata, X. Z. Yu, T. Hara, D. Morikawa, N. Kanazawa, K. Kimoto, S. Ishiwata, Y. Matsui, and Y. Tokura, Nat. Nanotechnol. {\bf 8}, 723-728 (2013).}





%
\bibitem{Science_349_234_2015}{Kirsten von Bergmann, Science {\bf 349}, 234-235 (2015).}
%


%
\bibitem{PhysRevB_85_174416_2012}{Y. Tchoe and J. H. Han, Phys. Rev. B {\bf 85}, 174416 (2012).}
%
\bibitem{ApplPhysLett_102_222405_2013}{S-Z Lin, C. Reichhardt, and A. Saxena, Appl. Phys. Lett. {\bf 102}, 222405 (2013).}
%
\bibitem{PhysRevB_88_184422_2013}{S. Rohart and A. Thiaville, Phys. Rev. B {\bf 88}, 184422 (2013).}
%
\bibitem{NatureNanoTechnology_8_742_2013}{J. Iwasaki, M. Mochizuki, and N. Nagaosa, Nat. Nanotechnol. {\bf 8}, 742-747 (2013).}
%
\bibitem{Nat_Commun_5_4652_2014}{Y. Zhou and M. Ezawa, Nat. Commun. {\bf 5}, 4652 (2014).}
%
\bibitem{PhysRevLett_114_177203_2015}{N. Romming, A. Kubetzka, C. Hanneken, K. von Bergmann, and R. Wiesendanger, Phys. Rev. Lett. {\bf 114}, 177203 (2015).}
%
\bibitem{JPhysD_ApplPhys_48_115004_2015}{Jinjun Ding, Xiaofei Yang, and Tao Zhu, J. Phys. D: Appl. Phys. {\bf 48}, 115004 (2015).}
%
\bibitem{PhysRevLett_110_167201_2013}{L. Sun, R. X. Cao, B. F. Miao, Z. Feng, B. You, D. Wu, W. Zhang, An Hu, and H. F. Ding, Phys. Rev. Lett. {\bf 110}, 167201 (2013).}
%
\bibitem{Scientific_Reports_5_17137_2015}{M. Beg, R. Carey, W. Wang, D. Cort\'{e}s-Ortu\~{n}o , M. Vousden, M.-A. Bisotti, M. Albert, D. Chernyshenko, O. Hovorka, R. L. Stamps, and H. Fangohr, Sci. Rep. {\bf 5}, 17137 (2015).}
%
\bibitem{AIP_Advances_5_047141_2015}{C. P. Chui, Fusheng Ma, and Yan Zhou, AIP Advances {\bf 5}, 047141 (2015).}
%
\bibitem{PhysRevB_93_024415_2016}{X. Zhang, Y. Zhou, and M. Ezawa, Phys. Rev. B {\bf 93}, 024415 (2016).}







%
\bibitem{Science_330_1648_2010}{F. Jonietz, S. M\"{u}hlbauer, C. Pfleiderer, A. Neubauer, W. M\"{u}nzer, A. Bauer, T. Adams, R. Georgii, P. B\"{o}ni, R. A. Duine, K. Everschor, M. Garst, and  A. Rosch, Science {\bf 330}, 1648-1651 (2010).}
%
\bibitem{Nat_Commun_3_988_2012}{X. Z. Yu, N. Kanazawa, W. Z. Zhang, T. Nagai, T. Hara, K. Kimoto, Y. Matsui, Y. Onose, and Y. Tokura, Nat. Commun. {\bf 3}, 988 (2012).}
%
\bibitem{Nat_Commun_4_1463_2013}{J. Iwasaki, M. Mochizuki, and N. Nagaosa, Nat. Commun. {\bf 4}, 1463 (2013).}
%
\bibitem{PhysRevLett_110_207202_2013}{Shi-Zeng Lin, C. Reichhardt, C. D. Batista, and A. Saxena, Phys. Rev. Lett. {\bf 110}, 207202 (2013).}
%
\bibitem{PhysRevB_89_064412_2014}{Junichi Iwasaki, Aron J. Beekman, and Naoto Nagaosa, Phys. Rev. B {\bf 89}, 064412 (2014).}
%
\bibitem{PhysRevB_89_064425_2014}{M. E. Knoester, J. Sinova, and R. A. Duine, Phys. Rev. B {\bf 89}, 064425 (2014).}
%
\bibitem{Scientific_Reports_5_10620_2015}{I. Purnama, W. L. Gan, D. W. Wong, and W. S. Lew, Sci. Rep. {\bf 5}, 10620 (2015).}


\bibitem{NaturePhysics_13_162_2017}{W. Jiang, X. Zhang, G. Yu, W. Zhang, X. Wang, M. B. Jungfleisch, J. E. Pearson, X. Cheng, O. Heinonen, K. L. Wang, Y. Zhou, A. Hoffmann, and S. G. E. te Velthuis, Nature Phys. {\bf 13}, 162-169 (2017).}







\bibitem{PhysRevLett_30_230_1973}{A. A. Thiele, Phys. Rev. Lett. {\bf 30}, 230-233 (1973).}
%
\bibitem{PhysRevB_14_3130_1976}{A. A. Thiele, Phys. Rev. B {\bf 14}, 3130-3165 (1976).}
%
\bibitem{NaturePhysics_8_301_2012}{T. Schulz, R. Ritz, A. Bauer, M. Halder, M. Wagner, C. Franz, C. Pfleiderer, K. Everschor, M. Garst, and A. Rosch, Nature Phys. {\bf 8}, 301-304 (2012).}




\bibitem{PhysRevLett_108_017601_2012}{Masahito Mochizuki, Phys. Rev. Lett. {\bf 108}, 017601 (2012).}
%
\bibitem{Nanotechnology_26_225701_2015}{X. Zhang, M. Ezawa, D. Xiao, G. P. Zhao, Y. Liu, and Y. Zhou, Nanotechnology {\bf 26}, 225701 (2015).}






\bibitem{PhysRevLett_111_067203_2013}{Lingyao Kong and Jiadong Zang, Phys. Rev. Lett. {\bf 111}, 067203 (2013).}
%
\bibitem{PhysRevLett_112_187203_2014}{S-Z Lin, C. D. Batista, C. Reichhardt, and A. Saxena, Phys. Rev. Lett. {\bf 112}, 187203 (2014).}



\bibitem{Scientific_Reports_6_24795_2016}{X. Zhang, Y. Zhou, and M. Ezawa, Sci. Rep. {\bf 6}, 24795 (2016).}
%
\bibitem{PhysRevB_96_060406_2017}{B. G\"{o}bel, A. Mook, J. Henk, and I. Mertig, Phys. Rev. B {\bf 96}, 060406(R) (2017).}


%
\bibitem{PhysRevB_89_054434_2014}{R. L. Silva, L. D. Secchin, W. A. Moura-Melo, A. R. Pereira, and R. L. Stamps, Phys. Rev. B {\bf 89}, 054434 (2014).}
%
\bibitem{PhysRevB_91_054410_2015}{Jan M\"{u}ller and Achim Rosch, Phys. Rev. B {\bf 91}, 054410 (2015).}
%




%
\bibitem{Ion_Irradiation_1998}{C. Chappert, H. Bernas, J. Ferr\'{e}, V. Kottler, J.-P. Jamet, Y. Chen, E. Cambril, T. Devolder, F. Rousseaux, V. Mathet, H. Launois, Science {\bf 280}, 1919-1922 (1998).}
%
\bibitem{Review_Implantation_2008} {J. Fassbender and J. McCord, J. Magn. Magn. Mater. {\bf 320}, 579-596 (2008).}
%
%
\bibitem{ApplPhysLett_101_252402_2012}{D. Toscano, S. A. Leonel, P. Z. Coura, F. Sato, R. A. Dias, and B. V. Costa, Appl. Phys. Lett. {\bf 101}, 252402 (2012).}
%
\bibitem{JMagnMagnMater_324_3083_2012}{J. H. Silva, D. Toscano, F. Sato, P. Z. Coura, B. V. Costa, and S. A. Leonel, J. Magn. Magn. Mater. {\bf 324}, 3083-3086 (2012).}

\bibitem{JMagnMagnMater_419_37_2016}{D. Toscano, S. A. Leonel, P. Z. Coura, F. Sato, B. V. Costa, M. V\'{a}zquez, J. Magn. Magn. Mater. {\bf 419}, 37-42 (2016).}

\bibitem{JApplPhys_116_163901_2014}{D. M. Burn and D. Atkinson, J. Appl. Phys. {\bf 116}, 163901 (2014).}
%
\bibitem{PhysRevApplied_3_034008_2015}{M. J. Benitez, M. A. Basith, R. J. Lamb, D. McGrouther, S. McFadzean, D. A. MacLaren, A. Hrabec, C. H. Marrows, S. McVitie, Phys. Rev. Applied {\bf 3}, 034008 (2015).}
%
\bibitem{ApplPhysLett_109_042406_2016}{M. V. Sapozhnikov, S. N. Vdovichev, O. L. Ermolaeva, N. S. Gusev, A. A. Fraerman, S. A. Gusev, and Yu. V. Petrov, Appl. Phys. Lett. {\bf 109}, 042406 (2016).}




\bibitem{PhysRevLett_115_267210_2015}{H. Yang, A. Thiaville, S. Rohart,A. Fert, and M. Chshiev, Phys. Rev. Lett. {\bf 115}, 267210 (2015).}
%
\bibitem{Nano_Lett_17_2703_2017}{W. Legrand, D. Maccariello, N. Reyren, K. Garcia, C. Moutafis,C. Moreau-Luchaire, S. Collin, K. Bouzehouane, V. Cros, and A. Fert, Nano Lett. {\bf 17}, 2703-2712 (2017).}



%
\bibitem{IEEE_51_1500204_2015} {H. T. Fook, W. L. Gan, I. Purnama, and W. S. Lew, IEEE Trans. Magn. {\bf 51}, 1500204 (2015).}
%
\bibitem{Scientific_Reports_7_45330_2017}{P. Lai, G. P. Zhao, H. Tang, N. Ran, S. Q. Wu, J. Xia, X. Zhang, and Y. Zhou, Sci. Rep. {\bf 7}, 45330 (2017).}
%






%



%



%

%

%
%
\bibitem{JPhys_CondensMatter_25_076005_2013}{Ye-Hua Liu and You-Quan Li, J. Phys.: Condens. Matter. {\bf 25}, 076005 (2013).}
%

\bibitem{PhysRevB_87_214419_2013}{S-Z Lin, C. Reichhardt, C. D. Batista, and A. Saxena, Phys. Rev. B {\bf 87}, 214419 (2013).}




\bibitem{JAP_109_076104_2011} {D. Toscano, S. A. Leonel, R. A. Dias, P. Z. Coura, and B. V. Costa, J. Appl. Phys. {\bf 109}, 076104 (2011).}
%
\bibitem{JAP_114_013907_2013} {V. A. Ferreira, D. Toscano, S. A. Leonel, P. Z. Coura, R. A. Dias, and F. Sato, J. Appl. Phys. {\bf 114}, 013907 (2013).}


\bibitem{PhysRevB_96_214403_2017}{D. Stosic, T. B. Ludermir, and M. V. Milosevi\'{c}, Phys. Rev. B {\bf 96}, 214403 (2017).}





%
%
%
%
%


%

%
\bibitem{Scientific_Reports_4_6784_2014}{R. Tomasello, E. Martinez, R. Zivieri, L. Torres, M. Carpentieri, and G. Finocchio, Sci. Rep. {\bf 4}, 6784 (2014).}

\end{thebibliography}
\end{document}